\documentclass[11pt]{JHEP3}

\usepackage{mathpazo}
\usepackage{amsmath}
\usepackage{amssymb}
\usepackage{graphicx}
\usepackage{cite}

\newcommand{\be}{\begin{eqnarray}}
\newcommand{\ee}{\end{eqnarray}}
\newcommand{\nn}{\nonumber}
\newcommand{\bn}{\begin{enumerate}}
\newcommand{\en}{\end{enumerate}}

\def\IC{\mathbb{C}}

\def\IR{\mathbb{R}}
\def\IZ{\mathbb{Z}}

\def\CB{{\cal B}}

\def\CL{{\cal L}}

\def\CN{{\cal N}}
\def\CO{{\cal O}}

\def\a{\alpha}
\def\b{\beta}
\def\g{\gamma}
\def\d{\delta}
\def\e{\epsilon}

\def\z{\zeta}
\def\th{\theta}

\def\k{\kappa}
\def\l{\lambda}

\def\s{\sigma}

\def\t{\tau}

\def\vph{\varphi}
\def\w{\omega}
\def\G{\Gamma}
\def\D{\Delta}

\def\S{\Sigma}

\def\Om{\Omega}

\def\half{\frac{1}{2}}
\def\thalf{{\textstyle \frac{1}{2}}}
\def\imp{\Rightarrow}

\def\goto{\rightarrow}

\def\grad{\nabla}

\def\p{\partial}
\def\tr{{\rm tr}}

\def\da{{\dot{\a}}}

\def\jmath{{j}}
\def\Lie{{\mathfrak{L}}}

\title{Schr\"odinger invariant solutions of M-theory
with Enhanced Supersymmetry}

\author{Jaehoon Jeong$^a$, Hee-Cheol Kim$^b$, Sangmin Lee$^{c,d}$,
Eoin \'O Colg\'ain$^d$, Hossein Yavartanoo$^d$
\\
\\
$^a$Department of Physics, College of Science, Yonsei University,
Seoul 120-749, Korea
\\
$^b$Department of Physics and Astronomy, Seoul National University,
Seoul 151-747, Korea
\\
$^c$Center for Quantum Space-time, Sogang University,
Seoul 121-742, Korea
\\
$^d$Korea Institute for Advanced Study, Seoul 130-722, Korea
\\
\\
}
\abstract{  
We find the most general solution of 11-dimensional supergravity
compatible with $\CN=2$ super-Schr\"odinger
symmetry with six supercharges and $SU(2)\times SU(2) \times U(1) \times \IZ_2$
global symmetry. It can be viewed as a one-parameter extension
of a recently constructed solution by Ooguri and Park. 
Our original motivation was to find the gravity dual of the
non-relativistic ABJM theory. But, our analysis shows that no such solution
exists within the reach of our assumptions. 
We discuss possible reasons for the non-existence of the desired solution.
We also uplift a super-Schr\"odinger solution in IIB supergravity 
of Donos and Gauntlett to 11-dimension and comment on its properties.

}


\preprint{KIAS-P09054}

\begin{document}

%\newpage

\section{Introduction}

The AdS/CFT correspondence in its various guises is now more than a decade old. {}From its original incarnation connecting $\mathcal{N}=4$ super Yang-Mills theory and near horizon geometry of D3-branes, via less supersymmetric models closer in nature to QCD, the conjectured AdS/CFT has passed numerous non-trivial hurdles, thus ensuring its place as one of the cornerstones of the string theory literature. 

Emboldened by such successes, physicists recently have shifted tack to applying the AdS/CFT to model conformal quantum mechanical condensed matter systems with non-relativistic (NR) symmetry. In this setting,  \cite{son,ba-mc} initiated a flurry of excitement in a NR version of the AdS/CFT by proposing a 
gravity background whose isometry group is the so-called Schr\"odinger group  
with dynamical exponent $z$. We will focus on the $z=2$ case, 
where the Schr\"odinger group consists of 
space and time translations, Galilean boosts, a scale transformation 
and a special conformal transformation.  

To study whether and how the NR-AdS/CFT works, it would be 
desirable to have a concrete example of a dual pair with 
a large amount of supersymmetry, {\it i.e.}, a NR analog of $\CN=4$ super-Yang-Mills 
and $AdS_5\times S^5$. 
A notable example in this regard is the ``non-relativistic mass deformed 
ABJM theory'' (NR-ABJM) constructed recently in \cite{naka, LLL} 
based on the $(2+1)$-dimensional $\CN=6$ Chern-Simons matter theory of 
Aharony, Bergman, Jafferis and Maldacena \cite{abjm}.
% 
%which has 12 Poincar\'e supercharges, 
%conformal supercharges with an $SO(6) \simeq SU(4)$ $R$-symmetry. 
%The mass deformation \cite{} preserves all Poincar\'e supercharges, 
%breaks all conformal supercharges and reduces the $R$-symmetry to 
%$SU(2)_1 \times SU(2)_2 \times U(1)_R$. 
%The usual non-relativistic limit keeps all particles and removes
%
The NR-ABJM theory has global symmetry group 
$U(1)_B \times SU(2)_1 \times SU(2)_2 \times U(1)_R \times \mathbb{Z}_2$ 
and 14 supersymmetries. 
%If a dual supergravity description exists, the high degree of symmetry 
%would make this an ideal place to test the NR-AdS/CFT duality.

The original ABJM theory at Chern-Simons level $(k,-k)$ 
describes multiple M2-branes probing the orbifold $\IC^4/\IZ_k$ 
in the transverse direction; the gravity dual is $AdS_4 \times S^7/\IZ_k$.\footnote{We will set $k=1$ for most of our discussion, although generalization 
for arbitrary $k$ is straightforward.} 
One may turn on an anti-self-dual four-form flux in $\IC^4$, 
which polarizes M2-branes into M5-branes \cite{bena1,BW,LLM}.
%\footnote{The extension to general level $k$ recently appeared in \cite{kumar}.}.  
This corresponds to the mass deformation of the ABJM theory \cite{mdABJM, gomis} 
with the most symmetric (classical) vacuum having the global symmetry $SU(2) \times SU(2) \times U(1) \times \mathbb{Z}_2$ and 12 Poincar\'e supersymmetries. 

In the course of taking the non-relativistic limit, the internal symmetry 
of the vacuum remains unchanged, while the space-time symmetry 
mutates into the Schr\"odinger symmetry. At the same time, 
the supersymmetry is enhanced from 12 to 14 supercharges. The latter may be 
divided into the sum of 2 dynamical, 2 kinematical and 2 conformal supercharges constituting the $\mathcal{N}=2$ super-Schr\"{o}dinger algebra \cite{dh,min,gala}, as well as 8 additional ``spectator" supercharges.

If the classical analysis of the field theory vacuum structure may be transplanted directly to the the supergravity setting, the gravity dual 
of the NR-ABJM theory could simply be found 
by taking a suitable ``non-relativistic limit" of the solution of \cite{BW,LLM}. 
However, as we will discuss below, there are some conceptual 
and technical difficulties for such an operation, 
which leads us to pursue an alternative approach. 
 
We start by constructing an ansatz for 11-dimensional supergravity 
that is compatible with all the global symmetry and Schr\"{o}dinger symmetry 
of the NR-ABJM theory, and proceed to analyze the Killing spinor equations. 
We succeed in finding the most general solution 
with 6 supercharges forming the $\CN=2$ super-Schr\"{o}dinger algebra.
The solution takes a simple, explicit form and includes 
two free parameters $(b,c)$. Setting $b=0$, we recover the one-parameter 
family of solutions previously found by Ooguri and Park \cite{oop}, 
where the result was obtained by deforming some known $\CN=1$ $AdS_5$ solutions in M-theory \cite{gaunt3}\footnote{Earlier work on non-relativistic deformations of this supersymmetric family appeared in \cite{eoin1}.}. 
%rather than a general ansatz or a deformation of $AdS_4$ solutions. 

However, bearing in mind the original goal of realizing the 8 additional supercharges, 
we are forced to conclude that, within the reach of our assumptions, 
the desired solution does not exist. 
We will list several possible explanations for the failure, 
but the discussion will not be conclusive.\footnote{ 
See \cite{rey} for a possibly related discussion.} 

We use standard methods for solving the Killing spinor equations, 
namely, spinorial Lie derivatives and G-structure. 
The methods may be easily adapted to generate more super-Schr\"{o}dinger solutions, but with the lengthy analysis involved, we confine ourselves to this one example. Instead, to illustrate how to compare with previously known $\CN=2$ super-Schr\"odinger 
solutions \cite{bobev,Donos:2009zf}, we perform T-duality on the IIB solution of \cite{Donos:2009zf} 
to obtain a new solution in M-theory containing an $S^2\times T^2$ component 
in the internal space. 

The rest of this paper is organized as follows. In section 2, we review some relevant features of both NR-ABJM theory and super-Schr\"odinger symmetry. We also exhibit our ansatz compatible with the expected global symmetries. In section 3, 
after a brief introduction to our methods, we present the solution, 
explain its main features, compare it with the solution of \cite{oop}. 
 Section 4 contains the details of solving the Killing spinor equations. In section 5, we uplift a IIB solution of \cite{Donos:2009zf} 
to M-theory and comment on its properties. 
We conclude in section 6 with a discussion on possible reasons 
why the gravity dual of NR-ABJM does not exist within reach of our assumptions.

\section{Motivation and Setup}

\subsection{Non-relativistic ABJM theory and BW/LLM solution}

\paragraph{A brief review of NR-ABJM}

The ABJM theory is an $\CN=6$ supersymmetric Chern-Simons-matter
theory with $U(N)\times U(N)$ gauge group
with Chern-Simons levels $(k,-k)$. The matter fields consist of 
bi-fundamental scalars $\Phi^A$ and fermions $\Psi_A$,
which transform under the $SU(4)\simeq SO(6)$ $R$-symmetry group
as ${\bf 4}$ and $\bar{\bf 4}$, respectively.

The theory is dual to M-theory on $AdS_4 \times S^7/\IZ_k$.
Regarding $S^7$ as a circle fibration over $\mathbb{CP}^3$,
the $\IZ_k$ acts on the the fiber. In other words,
the $\IZ_k$ action breaks the $SO(8)$ symmetry
of $S^7$ to $U(1)_B \times SU(4)$.
It is sometimes useful to take the $U(1)_B$ direction
to be the M-theory circle and consider IIA theory
on $AdS_4\times \mathbb{CP}^3$ with fluxes turned on.
In the field theory the $U(1)_B$ generator
counts the total number of bosons and fermions.

The non-relativistic ABJM theory with 14 supercharges (NR-ABJM) 
\cite{naka, LLL} can be obtained in two steps.
First, one performs a mass deformation \cite{mdABJM, gomis} 
which gives the same mass to all matter fields
(up to signs for fermions) and breaks the $SU(4)$ $R$-symmetry
into $SU(2)_1 \times SU(2)_2 \times U(1)_R$.
Second, one takes the usual non-relativistic limit for massive fields.
The Lagrangian of the resulting theory is as follows:
\be
\CL = \frac{k}{4\pi} \left( \CL_{\rm CS} + \CL_{\rm kin} + \CL_{\rm bos} + \CL_{\rm int1}+ \CL_{\rm int2} \right) \,,
\ee
where
\be
\CL_{\rm CS} &=& \e^{mnp} \tr\left[A_m \partial_n A_p -\tfrac{2i}{3} A_m A_n A_p -\tilde{A}_m \partial_n \tilde{A}_p +\tfrac{2i}{3} \tilde{A}_m \tilde{A}_n \tilde{A}_p \right]\,,
\nn \\
\CL_{\rm kin} &=& \tr\left[ \bar{\phi}_A (iD_t) \phi^A
-(D_i\bar{\phi}_A)(D_i\phi^A)\right]
\nn \\
&& + \tr\left[ \bar{\psi}^A (iD_t) \psi_A
+\bar{\psi}^a (D_i^2\psi_a -F_{12}\psi_a + \psi_a \overline{F}_{12} )
+\bar{\psi}^{\dot{a}} (D_i^2\psi_{\dot{a}} +F_{12}\psi_{\dot{a}} - \psi_{\dot{a}} \overline{F}_{12} ) \right]\,,
\nn \\
\CL_{\rm bos} &=& \thalf \tr\left[ \phi^a\bar{\phi}_{[a}\phi^b\bar{\phi}_{b]} -
\phi^{\dot{a}}\bar{\phi}_{[{\dot{a}}}\phi^{\dot{b}}\bar{\phi}_{{\dot{b}}]} \right] \,,
\nn \\
\CL_{\rm int1} &=& \tfrac{1}{4} \tr \left[ (\bar{\phi}_a \phi^a + \bar{\phi}_{\dot{a}} \phi^{\dot{a}})(\bar{\psi}^b \psi_b - \bar{\psi}^{\dot{b}} \psi_{\dot{b}} )  + (\phi^a\bar{\phi}_a  + \phi^{\dot{a}}\bar{\phi}_{\dot{a}} )(\psi_b\bar{\psi}^b  -  \psi_{\dot{b}}\bar{\psi}^{\dot{b}} )\right]
\nn \\
&& + \thalf \tr\left[ -\phi^a\bar{\phi}_b \psi_a \bar{\psi}^b +
\phi^{\dot{a}}\bar{\phi}_{\dot{b}} \psi_{\dot{a}} \bar{\psi}^{\dot{b}}
-\bar{\phi}_a \phi^b \bar{\psi}^a \psi_b  +
\bar{\phi}_{\dot{a}}\phi^{\dot{b}}\bar{\psi}^{\dot{a}} \psi_{\dot{b}}\right] \,,
\nn \\
\CL_{\rm int2} &=& -\thalf \tr\left[
\e^{ab}\e^{\dot{c}\dot{d}}
(\bar{\phi}_a \psi_b  \bar{\phi}_{\dot{c}} \psi_{\dot{d}}
+  \bar{\phi}_a \psi_{\dot{c}} \bar{\phi}_{\dot{d}} \psi_b )
+\e_{ab}\e_{\dot{c}\dot{d}} (
 \phi^a  \bar{\psi}^b \phi^{\dot{c}} \bar{\psi}^{\dot{d}}
+ \phi^a \bar{\psi}^{\dot{c}} \phi^{\dot{d}} \bar{\psi}^b )
\right]  \,.
\label{NR-ABJM-Lag}
\ee
We are mainly following the notations of \cite{naka} with some minor changes.
The $(a,b\,;\, \dot{a}, \dot{b})$ indices denote doublets of
$SU(2)_1 \times SU(2)_2$.

The Lagrangian (\ref{NR-ABJM-Lag}) is invariant under the scaling
\be
(t,x \; ; \; \phi, \psi) \;\; \goto \;\;
(\l^{-2} t, \l^{-1} x \; ; \; \l \phi , \l \psi) \,.
\ee
As in the free Schr\"odinger field theory,
this scaling symmetry can be extended to the full Schr\"odinger algebra 
which also includes a non-relativistic special conformal symmetry generator $K$.
%We will give a self-contained review of the Schr\"odinger
%algebra shortly.

As for the supersymmetry, all 12 Poincar\'e supercharges
of the ABJM theory survive the mass deformation as well as the non-relativistic limit.
Four of them are singlets under $SU(2)_1 \times SU(2)_2$.
Two of them $(Q,\bar{Q})$, which anti-commute to give the Hamiltonian $H$,
are called dynamical. The other two $(q,\bar{q})$ which
anti-commute to give the $U(1)_B$ generator are called kinematical.
These supercharges transform non-trivially under the Schr\"odinger algebra. 
In particular, the commutators between $K$ and $(Q,\bar{Q})$
require that an additional pair of supercharges $(S,\bar{S})$,
called conformal supercharges, should exist.
These six supercharges together with the Schr\"odinger generators 
form the so-called $\CN=2$ super-Schr\"odinger algebra 
as we will discuss in more detail below. 

The remaining eight supercharges $\{q_{a\dot{a}},\bar{q}^{a\dot{a}}\}$, which we call {\it spectators},
commute with all Schr\"odinger generators except the rotation, and transform in $({\bf 2}, {\bf 2})$ of $SU(2)_1 \times SU(2)_2$.
In summary, the NR-ABJM theory has the global symmetry group
\[
U(1)_B \times SU(2)_1 \times SU(2)_2 \times U(1)_R \times \IZ_2  \,,
\]
where the $\IZ_2$ interchanges the two $SU(2)$ factors,
and contains 14 supercharges.

\paragraph{BW/LLM solution and subtleties with the NR limit}

The gravity dual of the ABJM theory is AdS$_4\times S^7/\IZ_k$.
To find the gravity dual of the NR-ABJM theory,
a naive approach would be to carry over
the mass deformation and the non-relativistic limit to the gravity side.
But, a moment's thought reveals difficulties in such an attempt.

The gravity dual of the mass deformed theory
was obtained some time ago by Bena and Warner \cite{BW} (BW)
and reproduced later by Lin, Lunin and Maldacena \cite{LLM} (LLM);
see appendix \ref{bw-llm} for a short summary
of the BW/LLM solution. Bena-Warner begins with a collection of M2-branes
and turns on the four-form flux in the transverse directions.
The flux breaks the $SO(8)$ $R$-symmetry to $SO(4)\times SO(4)$
and polarizes the M2-branes into M5-branes,
which wrap the two three-spheres that are orbits of the $SO(4)$ groups.

There exists a gravity solution for each distinct configuration of
polarized M5-branes. Remarkably, the supergravity equations boil down
to a linear equation.
%, so the enumeration of different solutions become
%straightforward.
%
In the language of LLM, the smooth solutions
are in one-to-one correspondence with Young tableaux whose
total number of boxes $N$ are the same as the number of M2-branes
before polarization. It is widely believed that the polarized M5-branes correspond 
to ``fuzzy three-sphere'' configurations of the ABJM theory,
although an exact match at the quantum level still remains an open problem \cite{gomis}.

Note that the non-relativistic limit of the mass deformed ABJM theory was taken
without taking the polarization effects into account.
It would correspond to a BW/LLM solution with no polarized M5-branes.
Such a solution was written down in \cite{BW}, but was found to exhibit 
a naked singularity. The LLM dictionary makes it clear that 
the singularity is unavoidable. 

%It is conceivable that the problem with singular BW/LLM solution
%may be avoided by a sort of near horizon limit away from the singularity.
Putting the singularity problem aside for a moment,
let us consider how to perform the non-relativistic limit
on the gravity side.
The NR-ABJM theory is non-trivial when there are non-zero number
of particles, which is proportional to the eigenvalue of the 
$U(1)_B$ generator, 
which in turn gets identified with the 
central element $M$ of the Schr\"odinger algebra.

Recall that the $U(1)_B$ generator acts on the circle fiber of $S^7$. 
On the other hand, in the geometric realization of the Schr\"odinger algebra 
to be reviewed in the next subsection, $M$ is identified with a light-cone momentum.
The situation is strongly reminiscent of the discrete light-cone quantization (DLCQ) procedure taken in the context of Schr\"odinger geometry in \cite{hrr,mmt,abm}.
A crucial difference is that in our case the light-cone momentum is taken along a direction
transverse to the M2-brane world-volume. The existence of this light-cone momentum also hinders attempts to obtain solutions via consistent truncation, as were performed in \cite{mmt,gaunt5,eoin2}. 

In principle, one could proceed as follows.
First, one modifies the BW/LLM solution
by adding the particle number $M$.
In the IIA picture, it amounts to turning on the flux counting the D0-brane charge.
Second, one makes the standard coordinate change of the DLCQ procedure:
\footnote{
See \cite{oop} for a closely related discussion.}
\be
&& \tilde{\phi} = \phi - \a t \,, \;\;\; \tilde{t} = t \,
\nn \\
\imp
&& \widetilde{H} \equiv i \p_{\tilde{t}} = i\p_t - \a (-i\p_\phi) \equiv H - \a M \,,
\;\;\;
\widetilde{M} \equiv - i \p_{\tilde{\phi}} = -i\p_\phi \equiv M \,.
\ee
With a suitably chosen constant $\a$ and an appropriate scaling limit,
the light-cone Hamiltonian is identified with the
Hamiltonian of the non-relativistic theory.
%\be
%H_{\rm non-rel} = \lim_{c\goto \infty} \left( H_{\rm rel} - (\mbox{rest mass}) \right)
%\ee
The gravity description is expected to be valid
for a large value of $M$.

Coming back to the BW/LLM solution,
it is conceivable that the scaling limit
of the DLCQ procedure may push away the singularity of the unpolarized solution,
so that the final non-relativistic solution
becomes free of any singularity. Whether such a phenomenon happens
could be tested only by a direct computation.
Unfortunately, we are hindered by a technical difficulty; 
it is not clear how to turn on the $M$ momentum
and obtain the fully back reacted supergravity solution, 
as the $U(1)_B$ circle is fibered non-trivially along the $\mathbb{CP}^3$ base.

We are thus led to an alternative approach. 
We will begin with the most general ansatz consistent
with the symmetries of the NR-ABJM theory
and look for a supergravity solution preserving the same amount of supersymmetry.
Before writing down the ansatz, we review the super-Schr\"odinger algebra 
in some detail.

\subsection{Super-Schr\"odinger symmetry}

%As usual, we wish to rely heavily on the large degree of symmetry of
%NR-ABJM.

\paragraph{Bosonic algebra in arbitrary dimensions}

%Throughout this note, we use $d$ for the number of spatial dimensions,
%and $i,j$ indices for $SO(d)$ vectors.

%\paragraph{``Poincar\'e frame"}
The Schr\"odinger algebra Sch$_d$ contains an $SO(2,1)$ subalgebra 
among the time-translation ($H$), dilatation ($D$) and special conformal 
($C$) generators. 
\be
\left[D, H \right] = +2 H\,,
\qquad
\left[D, C \right] = -2 C\,,
\qquad
\left[H, C \right] = - D\,,
\ee
as well as the $SO(d)$ subalgebra,
\be
[M^{ij}, M^{kl} ] =
+\d^{jk}M^{il}+\d^{il}M^{jk}-\d^{ik}M^{jl}-\d^{jl}M^{ik} \,.
\ee
The remaining generators are space-translations ($P^i$) and Galilean boosts ($G^i$). 
They are vectors under the $SO(d)$,
\be
[M^{ij}, P^k] = +\d^{jk} P^i-\d^{ik} P^j \,,
\qquad
[M^{ij}, G^k] = +\d^{jk} G^i-\d^{ik} G^j \,,
\ee
and satisfy the following commutation relations:
\be
[D, P^i] = +P^i \,,
\; && \;
[D, G^i] = -G^i \,,
\\
{}[H, P^i ] = 0 \,,
\qquad
[C, P^i ] = +G^i \,,
\; && \;
[H, G^i ] = -P^i \,,
\qquad
[C, G^i ] = 0 \,.
\ee
Finally, we have the central extension 
with the ``rest-mass'' or the particle number, 
\be
{}[P^i,G^j ]= -\d^{ij} M \,.
\ee
All the generators above are {\em anti}-Hermitian.

It is sometimes useful to introduce a Virasoro-like notation,
\be
L_0 \equiv \thalf D \,, \;\;
L_{-1} \equiv H \,, \;\;
L_{+1} \equiv C \,,
\;\;\;\;
P_{-1/2}^i \equiv P^i\,, \;\;
P_{+1/2}^i \equiv G^i\,,
\;\;\;\;
M_{0} \equiv M \,.
\ee
Then, the commutation relations can be compactly summarized as
\be
{}[L_m, L_n] = (m-n) L_{m+n}\,,
\;\;\;
[L_m, P_r^i ] = \left( \thalf m -r \right) P_{m+r}^i \,,
\;\;\;
[P_r^i , P_s^j ] = (r-s) \d^{ij} M_{r+s} \,.
\label{bos-poin-vira}
\ee
%Note that the usual Hermitian conjugation do not give $(L_m)^\dagger = L_{-m}\,$, etc.
%We could define a BPZ-like conjugation
%to get the desired relations.
%

\paragraph{Global frame.}
As explained in \cite{nish}, the operator-state map naturally introduces
the following recombination of generators:
\be
&&\widehat{L}_0 \equiv \half(-iH-iC) \,,
\;\;\;
\widehat{L}_{\pm 1} \equiv \half(-iH+iC\pm D) \,,
\nn \\
&&\widehat{P}_{\pm1/2}^i = \frac{1}{\sqrt{2}}(-i P^i \mp  G^i) \,,
\;\;\;
\widehat{M}_0 = -i M_0 \,.
\ee
The new generators also satisfy Virasoro-like commutation relations,
\be
{}[\widehat{L}_m, \widehat{L}_n] = (m-n) L_{m+n}\,,
\;\;
[\widehat{L}_m, \widehat{P}_r^i ] = \left( \thalf m -r \right) \widehat{P}_{m+r}^i \,,
\;\;
[\widehat{P}_r^i , \widehat{P}_s^j ] = (r-s) \d^{ij} \widehat{M}_{r+s} \,,
\label{bos-alg-final}
\ee
as well as the conjugation relations
\be
(\widehat{L}_m)^\dagger = L_{-m}\,,
\;\;\;
(\widehat{P}_r^i)^\dagger = P_{-r}^i \,,
\;\;\;
(\widehat{M}_0)^\dagger = \widehat{M}_0\,.
\ee

\paragraph{Geometric realization}

In \cite{son,ba-mc}, a $(d+3)$-dimensional Schr\"odinger-invariant metric was presented. In our convention, the metric takes the form
\be
\label{son-met}
ds^2 = -\frac{dt^2}{r^4} + \frac{2dtdv+d\vec{x}^2+dr^2}{r^2} \,.
\ee
The generators of the Schr\"odinger algebra are realized as Killing vectors
of this metric,
\be
&&L_{m} = -  t^{m+1} \partial_t - \thalf(m+1) t^m(r\partial_r + x^i\partial_i) + \textstyle{\frac{1}{4}}m(m+1) t^{m-1}(\vec{x}^2 + r^2) \partial_v  \,,
\nn \\
&&P^i_{r} = t^{r+1/2}\partial^i -(r+\thalf) t^{r-1/2}x^i\partial_v \,,
\;\;\;\;\;
M_{m} = t^m \partial_v \,,
\;\;\;\;\;
M_{ij} = x_i \p_j - x_j \p_i \,.
\ee
A global coordinate for the geometry (\ref{son-met}) was found in \cite{blau}.
It is related to the Poincar\'e coordinate by the following transformation,
\be
t= \tan T \,, \;\;\;\;
r = \frac{R}{\cos T} \,, \;\;\;\;
\vec{x} = \frac{\vec{X}}{\cos T} \,, \;\;\;\;
v = V - \thalf(R^2+\vec{X}^2)\tan T \,.
\ee
In the new coordinate, the metric reads
\be
ds^2 = -\frac{dT^2}{R^4} + \frac{2dTdV-(\vec{X}^2+R^2)dT^2+d\vec{X}^2+dR^2}{R^2} \,.
\ee
The global form of the Schr\"odinger generators get simplified in this coordinate,
\be
&&\widehat{L}_0 = \thalf(i \p_T) \,,
\;\;\;\;
\widehat{L}_{\pm 1} = \thalf e^{\pm 2iT}
\left[ i \p_T + i (\vec{X}^2+R^2)\p_V \mp \left(X^i \p_{X^i} + R \p_R \right) \right] \,,
\nn \\
&&\widehat{P}^i_{\pm 1/2} = \textstyle{\frac{1}{\sqrt{2}}} e^{\pm iT}
\left( -i \p_{X^i} \mp  X^i \p_V \right) \,,
\;\;\;\;
\widehat{M}_0 = -i \p_V \,.
\ee

\paragraph{Schr\"odinger algebra in $d=2$}

Let $J\equiv -i M^{12}$ be the $SO(2)$ rotation generator.
It is useful to combine other generators
according to their helicity ($J$-eigenvalue) defined by
\be
{}[J, \CO] = j \CO\,.
\ee
For example, $P_r \equiv P_r^1 + i P_r^2$ has $j=+1$ and
$\bar{P}_r \equiv P_r^1 - i P_r^2$
has $j=-1$. In the helicity basis,
the bosonic algebra can be rewritten as
\be
{}[L_m, L_n] = (m-n) L_{m+n}\,,
\;\;\;
[L_m, P_r ] = \left( \thalf m -r \right) P_{m+r} \,,
\;\;\;
[P_r , \bar{P}_s ] = 2 (r-s) M_{r+s} \,.
\label{bos-alg-2d}
\ee
In what follows, we will denote operators with non-negative
$j$ by unbarred operators $\CO$ and their hermitian conjugates
by barred operators $\bar{\CO}$.

\paragraph{Super-Schr\"odinger algebra in $d=2$}

\paragraph{$\CN=2$ super-Sch algebra}
This algebra was first introduced in \cite{min} 
in the context of Chern-Simons systems. 
The notation $\CN=2$ refers to the supersymmetry of the
relativistic parent theory. In the ``Poincar\'e frame",
it has kinematical $(q,\bar{q})$, dynamical
$(Q,\bar{Q})$ and conformal $(S,\bar{S})$ supercharges,  
and a $U(1)$ $R$-symmetry.

Let us jump directly to the Virasoro-like notation in which the commutation relations take the simplest form. The supercharges are denoted
by $q$, $Q_{-1/2}\equiv Q$, $Q_{+1/2}\equiv S$ and their conjugates.
They transform under the $SO(2,1)\times U(1)_J \times U(1)_R$
subalgebra as
%\be
%{}[L_m, Q_r ] = \left( \thalf m -r\right) Q_r  \,, \;\; & \;\;
%{}[J, Q_r ] = +\thalf Q_r \,, \;\; & \;\;
%{}[R, Q_r] = + Q_r \,,
%\\
%{}[L_m, q] = 0\,, \qquad\qquad \;\;\;\;\;  & \;\;
%{}[J, q] = +\thalf q \,, \;\; & \;\;
%{}[R, q] = - q \,.
%\ee
\be
{}[L_m, Q_r ] = \left( \thalf m -r\right) Q_r  \,, \;\;\;\;
{}[L_m, q] = 0\,,
\ee
and
\be
{}[J, Q_r ] = +\thalf Q_r \,, \;\;\;\;
{}[R, Q_r] = + Q_r \,, \;\;\;\;
{}[J, q] = +\thalf q \,, \;\;\;\;
{}[R, q] = - q \,.
\ee
Their commutators with $P_r$ give
\be
{}[\bar{P}_r , Q_s ] = (r-s) \bar{q}\,,
\qquad
{}[\bar{P}_r , q ] = 0\,.
\ee
Finally, the anti-commutators among supercharges give
\be
\label{n2sch3}
\{ \bar{Q}_r , Q_s \} = L_{r+s} + \thalf (r-s)\left(J -\textstyle{\frac{3}{2}} R \right) \,,
\qquad
\{q, Q_r \} = P_r \,,
\qquad
\{\bar{q}, q \} = 2M \,.
\ee
Note that $(L_m,Q_r, J-\frac{3}{2}R)$ form a closed sub-algebra,
called $OSp(2|1)$,
isomorphic to the usual $\CN=2$ superconformal algebra
in a chiral sector of RNS superstring world-sheet.

%\paragraph{$\CN > 2$ super-Sch algebras}
%The $\CN=3$ example of Nakayama et al (arXv:0811.2461),
%the $\CN=6$ NR-ABJM and
%even the $\CN=8$ NR-BLG contain the $\CN=2$ super-Schr\"odinger algebra.
%While the $\CN=2$ super-algebra can be understood
%as a DLCQ-like subalgebra of $SU(2,2|1)$,
%the other super-algebras realized by Chern-Simons
%theories cannot be understood as subalgebras of $SU(2,2|\CN)$.

\paragraph{$\CN=6$ super-Sch algebra}
The ABJM theory has an $SU(4)$ $R$-symmetry.
The mass deformation breaks it into $SU(2)_1
\times SU(2)_2 \times U(1)_R$.
The six supercharges participating in the $\CN=2$ subalgebra
are singlets of $SU(2)_1\times SU(2)_2$.
The additional eight supercharges, which we call {\it spectator} supercharges,
satisfy the following relations:
\be
&&{}[L_m, q_{a\dot{a}} ] = 0  \,, \;\;\;
{}[P_r , q_{a\dot{a}} ] = 0  = {}[\bar{P}_r , q_{a\dot{a}} ] \,,
\nn \\
&&{}\{Q_r , q_{a\dot{a}} \} = 0 =
{}\{\bar{Q}_r , q_{a\dot{a}} \}  \,, \;\;\;
{}\{q , q_{a\dot{a}} \} = 0  =
{}\{\bar{q} , q_{a\dot{a}} \}  \,,
\nn \\
&&{}[J, q_{a\dot{a}} ] = +\thalf q_{a\dot{a}} \,, \;\;\;
{}[R, q_{a\dot{a}} ] = 0 \,,
\nn \\
&&
{}[R^a{}_b , q_{c\dot{c}} ] = -\d^\a_\g q_{b\dot{c}} + \thalf \d^a_b q_{c\dot{c}} \,, \;\;\;\;
{}[R^{\dot{a}}{}_{\dot{b}} , q_{c\dot{c}} ] = -\d^a_c q_{\dot{c}\dot{b}} + \thalf \d^{\dot{a}}_{\dot{b}} q_{c\dot{c}} \,,
\nn \\
&&\left\{ \bar{q}^{a\dot{a}} , q_{b\dot{b}} \right\}
= \half \d^a_b \d^{\dot{a}}_{\dot{b}} M - \d^a_b R^{\dot{a}}{}_{\dot{b}} +\d^{\dot{a}}_{\dot{b}} R^a{}_b \,,
\ee
where $R^a{}_b$, $R^{\dot{a}}{}_{\dot{b}}$ are the $SU(2)$ generators defined by
\be
[R^a{}_b, R^c{}_d ] = \d^c_b R^a{}_d - \d^a_d R^c{}_b \,,
\;\;\;\;
(R^a{}_b)^\dagger = R^b{}_a \,.
\ee
The $\CN=2$ subalgebra (\ref{n2sch3}) still holds,
except that the generator $R$ is replaced by $\tilde{R}$.
In the field theory, the shift is partly due to an additional conserved quantity,
namely, the fermion number $\S$.
The shifted $R$-charge is related to the original one by
$\tilde{R} = (4/3)R - (2/3)\S$.
{}From the commutation relations, we see that 
the shift is needed to make $q_{a\dot{a}}$ 
neutral under $J-\frac{3}{2}\tilde{R}$, 
which should hold because $q_{a\dot{a}}$ commutes with $Q_r$. 
It is not clear how to realize $\S$ geometrically.
\begin{table}[t]
\begin{center}
\begin{tabular}{c|c|c|c|c}
  & $L_0$ & $J$   & $R$ & $\tilde{R}$  \\
  \hline
$Q$ & $+1$ & $+1/2$ & $+1$ & $+1$ \\
$S$ & $-1$ & $+1/2$ & $+1$ & $+1$ \\
$q$ & $0$ & $+1/2$ & $-1$ & $-1$ \\
$q_{\a\da}$ & $0$ & $+1/2$ & $0$ & $1/3$
\end{tabular}
\end{center}
\caption{$U(1)$ quantum numbers of supercharges.}
\label{twi-all}
\end{table}

%\paragraph{Geometric realization}

%%%%%%%%%%%%%%%%%%%%%%%%%%%%%%%%%%%%%%%%%

%\newpage
%\section{The Solution}

\subsection{Ansatz}

Recall the sequence of the $R$-symmetry breaking,
\be
SO(8) \supset U(1)_B \times SU(4)
\supset U(1)_B \times SU(2)_1 \times SU(2)_2 \times U(1)_R \,.
\label{r-symm}
\ee
%The central element $M$ of the schr\"odinger algebra, which
%counts the total number of bosons and fermions,
%corresponds to the $U(1)_B$ generator.
%
To see how these $R$-symmetries are realized geometrically,
consider $S^7$ as a warped product of two $S^3$'s,
and write down the metric as
\be
ds^2_{S^7} = d\a^2 + \cos^2\a \,d\Om_1^2 + \sin^2\a \,d\Om_2^2 \,.
\ee
We use the standard Euler-angle coordinates ($\th,\phi,\psi$) for each $S^3$:
\be
d\Om^2_i = \frac{1}{4} \left[ d\th_i^2 + \sin^2\th_i d\phi_i^2 + (d\psi_i- \cos\th_i d\phi_i )^2\right]
\;\;\;\;\; (i=1,2,\; \mbox{no sum}).
\ee
We choose the orientations of the 3-spheres such that
the $U(1)_R$ acts diagonally on $\psi_{1,2}$
and the $U(1)_B$ acts with an opposite relative sign.

Now, let us begin with AdS$_4 \times S^7/\IZ_k$ and
imagine taking the mass deformation and then the
non-relativistic limit. The procedure will change the
metric significantly, but the $R$-symmetries
(\ref{r-symm}) as well as the time and space translation
(in Poincar\'e patch) should be preserved throughout.
Moreover, the fibration structure of the $U(1)_B$ and $U(1)_R$ angles
over the two $S^2$'s should be maintained.

In what follows, we will use the following notations
\be
&&w = \thalf(\psi_1 + \psi_2) \,, \;\;\;
v = \thalf(\psi_1 - \psi_2) \,,
\\
%\;\;\; (\mbox{we will also use}\;\; v\equiv\psi_-, \;\;\; w\equiv\psi_+)
%\ee
%and defined
%\be
&&D w = d w - \thalf(\cos\th_1d\phi_1 + \cos\th_2 \phi_2) \,,
\\
&&D v = d v - \thalf(\cos\th_1d\phi_1 - \cos\th_2 \phi_2) \,,
\\
&&d\w_i^2 = \textstyle{\frac{1}{4}} (d\th_i^2 + \sin^2 \th_i d\phi_i^2) \,,
%\;\;\;
%J_i = \textstyle{\frac{1}{4}} \sin\th_i d\th_i \wedge d\phi_i \,.
\ee

\paragraph{Metric}
We can try to write down the most general ansatz
for the metric and the 4-form flux consistent with 
the Schr\"odinger symmetry, global symmetries 
as well as the fibration structure.
Building upon the Schr\"odinger-invariant metric of \cite{son,ba-mc},
\be
ds^2 = -\frac{dt^2}{r^4} + \frac{2dtd\psi +dr^2+d\vec{x}^2}{r^2} \,,
\ee
we propose our ansatz for the metric, 
\be
\label{met-ans1}
ds^2 &=& e^{2c_1} \left(-c_2 \frac{dt^2}{r^4} + \frac{2dt (Dv + c_3 D w)+dr^2+d\vec{x}^2}{r^2} + \frac{4}{9} e^{2h_2} (Dw)^2 \right)
\nn \\
&&+ e^{-4c_1}\left(  e^{-2h_2} d y^2+ \frac{4}{3} e^{2h_1} (e^{+2h_3} d\w_1^2 + e^{-2h_3} d\w_2^2) \right)  \,.
\ee
All the functions ($c_{1,2,3}$, $h_{0,1,2,3}$)
depend only on $y$, which is the only coordinate
not constrained by the continuous symmetries of the geometry.
We ``gauge-fixed" the reparametriztion invariance in $y$
by a particular choice of $g_{yy}$.
The numerical factors $4/9$ and $4/3$ are inserted
for later convenience.
The Schr\"odinger symmetry and $R$-symmetry allow for
two more terms in the metric,
\be
r^{-2} dt dy \,, \;\;\; Dw dy \,,
\ee
but both of them can be removed by shifting $v$ and $w$
by $y$-dependent functions.

\paragraph{Orthonormal frame}

The metric ansatz (\ref{met-ans1}) admits a natural orthonormal frame,
\be
&&e^+ = \frac{e^{2c_1}}{r^2} dt \,, \;\;\;
e^- = -\frac{c_2}{2r^2}dt + Dv + c_3 Dw  \,,
\nn \\
&&e^1 = \frac{e^{c_1}}{r} dx^1 \,, \;\;\;
e^2 = \frac{e^{c_1}}{r} dx^2 \,, \;\;\;
e^7 = \tfrac{2}{3} e^{c_1+h_2} Dw \,, \;\;\;
e^8 = \frac{e^{c_1}}{r} dr \,, \;\;\;
e^9 = e^{-2c_1-h_2} d y \,, \;\;\;
\nn \\
&&(e^3,e^4 \; ; \; e^5,e^6) = \frac{1}{\sqrt{3}} e^{-2c_1+h_1}
\left( e^{+h_3}(\s_1,\s_2) \; ; \; e^{-h_3}(\t_1,\t_2)\right) \,.
\label{ortho}
\ee
Here, $\s_A$, $\tau_A$ are invariant one forms of $S^3$'s.
See appendix A for our convention for Euler-angle coordinates.

\paragraph{Flux}
To write down the general ansatz for the 4-form flux,
we first collect all Schr\"odinger invariant $p$-forms in the ``external'' part of the metric:
\be
\{ e^{+128}, e^{+12}, e^{+8}, e^{+} \}
\ee
%\be
%&&A_4 = r^{-5} dr \wedge dt \wedge dx^1 \wedge dx^2 \,,
%\;\;\;
%B_3 = r^{-4} dt \wedge dx^1 \wedge dx^2 \,,
%\nn \\
%&&C_2 = r^{-3} dr \wedge dt   \,,
%\;\;\;
%D_1 = r^{-2} dt  \,.
%\ee
%({\bf Are we sure that this list is complete?})
Note that all the invariant $p$-forms contain $e^+$.
Combining these with invariant $p$-forms from the internal part,
we arrive at the ansatz with ten unknown functions,
\be
\label{flux-ans1}
F &=&
e^{-3c_1} e^{+8} \left[ e^{-2c_1} k_1 e^{12}+ e^{4c_1-2h_1}(e^{-2h_3}k_{4,1} e^{34}+e^{+2h_3}k_{4,2} e^{56})\right]
\nn \\
&&+e^{h_2} e^{+9} \left[ e^{-2c_1}k_2 e^{12}+ e^{4c_1-2h_1}(e^{-2h_3}k_{5,1} e^{34}+e^{+2h_3}k_{5,2} e^{56})\right]
\nn \\
&&+ e^{c_1} e^{97}\left[e^{-3c_1} k_3 e^{+8}+  e^{4c_1-2h_1}(e^{-2h_3}k_{6,1} e^{34}+e^{+2h_3}k_{6,2} e^{56}) \right]
\nn \\
&&+e^{8c_1-4h_1}  k_7e^{3456} \,.
\ee
Here, we are using the shorthand notation $e^{ab} = e^a \wedge e^b$, etc.
and assuming wedge products among differential forms.
We inserted compensating factors of metric coefficients so that
the Bianchi identity ($dF=0$) maintains the simple form,
\be
k_1' + 4k_2 &=& 0 \,,
\nn \\
k_{4,1}'+2 k_{5,1} - k_3 &=& 0 \,,
\nn \\
k_{4,2}'+2k_{5,2} - k_3 &=& 0 \,,
\nn \\
k_7' - (k_{6,1}+k_{6,2}) &=& 0\,.
\ee
There are three more terms allowed by the symmetries, $
\{ e^{+127}, e^{+347}, e^{+567} \}$,
but they are excluded by the Bianchi identity.

\paragraph{Parity symmetry}
There is a discrete $\IZ_2$ symmetry exchanging the two 2-spheres
which acts as a parity $y \goto -y$.
The unknown functions have the following parity eigenvalues,
\be
\mbox{Even} &:& c_1, c_2, h_1, h_2, k_1, (k_{4,1}+k_{4,2}), (k_{5,1}-k_{5,2}),
(k_{6,1}+k_{6,2}) \,.
\nn \\
\mbox{Odd} &:& c_3, h_3, k_2, k_3, (k_{4,1}-k_{4,2}), (k_{5,1}+k_{5,2}), (k_{6,1}-k_{6,2}), k_7 \,.
\ee

\section{Solution and a Sketch of the Computation}

Having written out the most general ansatz, %and imposed the Bianchis,
in this section we give a quick overview of our methods,
summarize the equations imposed on
the unknown functions in the ansatz,
write down the solution and discuss its properties.
The details of the computation will be postponed until the next section.

\subsection{Methods \label{KSE-meth}}

Supersymmetric solutions of M-theory satisfy the Killing spinor equation, 
\be
\d_\e \psi_M = \grad_M\e +\frac{1}{12\cdot 4!}F_{IJKL}
\left({F^{IJKL}}_M - 8 \d^{I}_{M} \G^{JKL} \right)\e = 0\,.
\ee 
See appendix A for our conventions for 11-dimensional supergravity. 
Our approach to the problem will hinge upon
two standard tools used for finding supersymmetric solutions,
namely, the spinorial Lie derivative and the G-structure.

To begin with, the Lie derivative of a spinor $\epsilon$ with respect to a Killing vector $K$ may be defined as in \cite{FigueroaO'Farrill:1999va}
\be
\Lie_K \e = K^m \nabla_m \e + \frac{1}{4} \left( \nabla_a K_b \right) \G^{ab} \e .
\ee
In general, the spinorial Lie derivative gives a geometric realization of the algebra,
\be
[K, Q_1 ] = Q_2  \;\;\; \Longleftrightarrow \Lie_K \e_{Q_1} = \e_{Q_2} \,.
\ee
{}From the metric ansatz (\ref{met-ans1}), one may then %press ahead and
write out the spinoral Lie derivatives associated to the various Killing directions. The Lie derivatives of the spinors, via the super Schr\"odinger algebra discussed in section 2, determine all coordinate dependence
other than the $y$-direction of the two dynamical supercharges $Q$. Once $Q$ are determined, the kinematical $q$ and conformal $S$ supercharges also may be worked out from the algebra.

Adopting the language of G-structures to M-theory was initiated in \cite{gaunt1,gaunt2}.
Assuming the existence of Killing spinors $\{ \e_i\}$,
one constructs the following differential forms
\be
K_{ij} &=& (\bar{\e_i} \G_{a} \e_j) e^{a} \,,
\label{vec-spin}
\\
\Omega_{ij} &=& \half (\bar{\e_i} \G_{ab} \e_j) e^{ab} \,,
\\
\S_{ij} &=& \frac{1}{5!} (\bar{\e_i} \G_{abcde} \e_j) e^{abcde}  \,.
\ee
%The Killing spinor equation implies
%\be
%\label{ext-K}
%dK &=& \frac{2}{3} i_\Om F + \frac{1}{3} i_{\S} \ast F \,,
%\\
%d\Om &=& i_K F \,,
%\label{ext-O}
%\\
%d\S &=& i_K \ast F - \Om \wedge F \,.
%\label{ext-S}
%\ee
The Killing spinor equations imply that $K_{ij}$ are
Killing vectors, so that (\ref{vec-spin}) becomes a geometric representation
of the algebra
\[
\{ Q_i , Q_j \} = K_{ij} \,.
\]
In addition, the KSE give a set of
algebraic and differential relations among $(K,\Om,\S)$.
These relations are equivalent to
the original KSE by construction,
but are often easier to solve and illuminate
the geometric structure more clearly.
For the purpose of this paper, however, it turns out to be
more straightforward to analyze the
KSE directly, while keeping in mind the lessons from \cite{gaunt1,gaunt2}.

We will demand that our ansatz admit
the six supercharges of $\CN=2$ super-Sch algebra.
The kinematical supercharges $(q,\bar{q})$
correspond to null Killing spinors studied in \cite{gaunt2},
whereas the dynamical supercharges $(Q,\bar{Q})$
correspond to time-like Killing spinors studied in \cite{gaunt1}.
To use the results of \cite{gaunt2} directly, we first focus on the real combination $\e=\thalf(q + \bar{q})$ which satisfies the two projection
conditions
\[
\G^{3456} \e = - \e \;\; (\mbox{singlet under $SU(2)_1 \times SU(2)_2$})\,,\;\;\;
\G^{+} \e = 0\,,
\]
and defines an $SU(7)$ structure explained in \cite{gaunt2}.
Restoring both components $(q,\bar{q})$ then defines an $SU(4)$ sub-structure of the Spin(7) structure. Having started by introducing an ansatz, making the G-structure manifest entails a small frame rotation from the original frame to the canonical G-structure frame.
Similarly, for $(Q,\bar{Q})$ we find an $SU(4)$ sub-structure
of the $SU(5)$ structure introduced in \cite{gaunt1}.
The conformal supercharges $(S,\bar{S})$ do not
yield any new information because they are related to $(Q,\bar{Q})$
by the conformal symmetry generator
and all bosonic symmetries are already built into our ansatz.

%There the Killing spinor $\e$ satisfies the expected projectors $\G^{1234} \e = \G^{3456} = \G^{5678} \e = - \e$ on the base eight-manifold with the projector $\G^{1357} \e = - \e$ being relaxed.

%The following steps then involve writing out the Killing spinor equation for $\e$, before sandwiching with $\bar{\e}$ to extract the non-zero components. Via this process, one derives a set of algebraic and differential relations that follows directly from the Killing spinor equation.
%We now turn our attention to these conditions and the resulting solution.

\subsection{Killing spinor equations: summary \label{KSE-sum}}

After a somewhat lengthy analysis to be presented in section \ref{detail}, the Killing spinor equations for the six supercharges
give rise to a number of coupled equations
for all the unknown variables. They may be divided into three blocks.

\bn

\item
Block A : The equations for $(c_1, h_1, h_2, h_3)$
decouple from all other variables.
\be
&& 4h_1' -h_2' = -c_1'(2h_1'+h_2')^2 e^{6c_1+2h_2} \,,
\label{a1x}
\\
&& 9c_1 ' = (9c_1'-4h_1' +h_2')e^{2h_2} \,,
\label{a2x}
\\
&&2h_1'+h_2' = 6(h_1'+h_3') e^{-6c_1 +2h_1-2h_2+2h_3} \,,
\label{a3x}
\\
\label{h1x}
&& h_3' \cosh(2h_3) = -h_1' \sinh(2h_3)
%\;\; \imp \;\; |\sinh(2f_2)| = \k_1 e^{-2h_1}
\label{h3x} 
\,.
\ee
The following auxiliary equations will also be useful,  
\be
&&\cos\z = e^{h_2}\,, \;\;\; 
\\
&&\sin\z = -\tfrac{1}{3} (2h_1'+h_2')e^{3c_1+2h_2} 
= \frac{1}{3c_1'}(-\z'\cos\z +2 e^{-3c_1}) \,.
\label{szx}
\ee

\item
Block B : With the solutions of Block A as an input,
we can solve the equations for $(c_3, k_1, k_2, k_3)$.
%In what follows, $\sin\z \equiv -\tfrac{1}{3} (2h_1'+h_2')e^{3c_1+2h_2}$.
\be
&&
k_2 = -k_3  \,,
\\
&&
k_1 = -\frac{6c_3}{\sin\z} e^{3c_1} \,,
\\
&& 3c_3' + k_1 e^{-6c_1} = 6\sin\z(c_3 \cosh(2h_3) -\sinh(2h_3) ) e^{3c_1-2h_1} \,,
\\
&&
3c_3'  = 2 \left( k_1 e^{-6c_1} \frac{h_1'-h_2'}{2h_1'+h_2'}  -k_3 e^{-3c_1} \sin\z \right) \,.
\ee

\item
Block C : The last metric component $c_2$ and all the remaining flux
components are determined algebraically by the solutions of Block A and Block B.
\be
&&
c_2 = \left( \tfrac{1}{4} k_1 e^{-3c_1} \right)^2 \,,
\\
&&
k_{4,1} =  -\tfrac{3}{2}(c_3+1)e^{3c_1}\sin\z
- \tfrac{1}{4} k_1(2e^{-6c_1+2h_1+2h_3}- e^{2h_2})  \,,
\\
&&
k_{4,2} =  -\tfrac{3}{2}(c_3-1)e^{3c_1}\sin\z
- \tfrac{1}{4} k_1(2e^{-6c_1+2h_1-2h_3}- e^{2h_2})  \,,
\\
&&
k_{5,1} = -\tfrac{3}{2}(c_3-1) e^{+4 h_2} \,,
\\
&&
k_{5,2} = -\tfrac{3}{2}(c_3+1) e^{-4 h_2} \,,
\\
&&
k_{6,1} = - \frac{h_1'+2h_2'+3h_3'}{3(h_1'+h_3')} e^{2h_2} \,,
\\
&&
k_{6,2} = - \frac{h_1'+2h_2'-3h_3'}{3(h_1'-h_3')} e^{2h_2} \,,
\\
&&
k_7 = 6c_1' e^{-6c_1+4h_1} \,.
\ee

\en
%Once we find a solution to the full Killing spinor equation,
%we can also check the Bianchi identity,
%Einstein equation and Maxwell equation.

\subsection{Solution}

Rather remarkably, the set of coupled equations listed above can be solved 
completely in a closed form. We first note that (\ref{h3x}) can be readily 
integrated to give 
\be
|\sinh(2h_3)| = \b e^{-2h_1} \,,
\label{f2-sol}
\ee
where $\b$ is an integration constant. 
But, for any non-vanishing $\b$, the metric is singular at $h_3=0$. To avoid the singularity, we are forced to set $\b=0$. Then, $h_3$ vanishes identically. 

Integrating the second half of (\ref{szx}), we find 
\be
\partial_y(\sin \zeta e^{3c_1}) = 2 \quad \Rightarrow \quad \sin\zeta e^{3c_1} = 2y\,.
\ee
In principle, another integration constant should be introduced here. 
But, a non-zero constant turns out to induce terms proportional to $(\log y)$ in $e^{2h_2}$ and $e^{6c_1}$, leading to a singular metric. So, we drop the constant.

Inserting the first half of (\ref{szx}) to the LHS of (\ref{h1x}) and integrating, 
we find
\be
e^{2h_1} = p-y^2 \,,
\ee
Multiplying each side of (\ref{a1x}) and (\ref{a3x}), eliminating $c_1'$ 
by using (\ref{a2x}) and integrating, we obtain the solution for $h_2$, 
\be
e^{2h_2} = \frac{-3y^4-2c y^3-6py^2+p^2}{(p-y^2)^2} \,,
\ee
Finally, solving (\ref{a3x}) for $e^{6c_1}$, we find 
\be
e^{6c_1} = \frac{(p-y^2)^2}{p+\thalf cy+y^2} \,.
\ee
Here, $c$ an $p$ are integral constants. By a rescaling of $y$ and $c$, 
as well as an overall rescaling of the whole metric, we can always set $p=1$. 
Written in this form, the metric components we have found 
are essentially identical to those of \cite{gaunt3}. 
The condition for global regularity of the solution discussed in \cite{gaunt3} 
can be carried over to our case; we require the following constraints on $c$ and $y$
\be
0 \leq c < 4 \,, \quad \quad  y_1 \leq y \leq y_2 \,,
\ee
where $y_1$ and $y_2$ are the two real roots of the equation $e^{2h_2} = \cos^2\zeta = 0$. In addition, we must choose the period of $w$ to be $2\pi$ to have a smooth geometry at both $y_1$ and $y_2$. The regular solutions with these conditions are $S^2$ fibrations over $S^2\times S^2$ \cite{gaunt3}.

It is straightforward to solve equations in Block B. For instance, by combining the second and third equations, we obtain
\be
c_3 =  \frac{4by}{3y(1-y^2)} \,.
\ee
with $b$ being an integration constant. Other equations in Block B are easier to solve. Block C equations then determine the remaining unknown functions algebraically. 

In summary, we have obtained the most general solution compatible with 
$\CN=2$ super-Schr\"odinger symmetry and the global symmetry discussed in section 2. The solution is parameterized by two real constants $b$ and $c$.
%\subsection{Solution}
%Restricting ourselves to the even parity solutions with $\mathbb{Z}_2$ symmetry between the two spheres, it is possible to solve these equations in a closed form, up to introduction of integration constants. Thus, imposing $h_3 = 0$, one may solve the block A equations above for ($c_1,h_1,h_2$) and one recovers the solution appearing in \cite{oop}. This is the only unique solution to this set of equations. One may then turn one's attention to Block B, which is also easily solved, producing expressions for ($c_3,k_1,k_2,k_3)$ . Finally, block C is purely alegebraic and is determined by the solutions of blocks A and B.
%\paragraph{Summary}
The final form of the solution may be most neatly captured in terms of two quadratic polynomials,
\be
g_1 = 1-y^2 \,, \;\;\;
g_2 = 1+\thalf cy+y^2 \,.
\ee
The metric components are
\be
&&e^{6c_1} = g_1^2g_2^{-1} \,,
\;\;\;\;\;
c_2 = b^2 g_1^{-2}g_2^{-1} \,,
\hskip 1cm
c_3 =  \tfrac{4}{3} b y g_1^{-2} \,,
\\
&&e^{2h_1} =  g_1 \,,
\hskip 0.8cm
e^{2h_2} = 1-4y^2e^{-6c_1} \,,
\hskip 0.5cm
e^{2h_3} = 1 \,,
\ee
and the flux components are
\be
&&k_1 = -4b g_2^{-1} \,,
\;\;\;\;\;\;\;
k_2 = -b g_2' g_2^{-2} \,,
\;\;\;\;\;\;\;
k_3 = b g_2' g_2^{-2} \,,
\\
&&k_{4,1} =  -3y + b(2g_1^{-1}-g_2^{-1})
\;\;\;\;\;
k_{4,2} =   +3y + b(2g_1^{-1}-g_2^{-1})  \,,
\\
&&k_{5,1} =   +\tfrac{3}{2} -2y b g_1^{-2}
\hskip 1.9cm
k_{5,2} =     -\tfrac{3}{2} -2y b g_1^{-2} \,,
\\
&&k_{6,1} = k_{6,2} = 1 - 4g_2 g_1^{-2} \,,
\hskip 1cm
k_7 = -4g_2'g_1^{-1} +2g_1'+3g_2' \,.
\ee

%\subsection{Discussion}

\paragraph{Ooguri-Park solution}

In hindsight, our solution can be 
regarded as a one-parameter generalization of the recently discovered 
solution of Ooguri and Park \cite{oop}. 
Their solution was obtained by a judicious deformation of 
a known $AdS_5$ solution in M-theory \cite{gaunt3}. 
It has two parameters $\beta$ and $c$. 
It preserves two supercharges for $\b \neq 0$ and 
six supercharges for $\b=0$. 

It is easy to compare our solution with the Ooguri-Park solution. 
To be consistent with normalization conventions,
we should relate our coordinates to theirs by
\be
t = -2x^+ \,, \;\;\;
v = \thalf x^- \,, \;\;\;
w = \thalf \psi \,,
\ee
and set $n=1$ in their solution, although it is easy to generalize
the comparison for arbitrary $n$. 
Then it is immediately clear that our solution with $b=0$ is 
identical to their solution with $\b=0$. 
Note that the norm of the `time' Killing vector $\p_t$ 
vanishes when $b=0$. 
In this sense, the $b\goto 0$ limit is not smooth; 

\paragraph{Non-existence of spectator supercharges}

Our original goal was to find the gravity dual of the NR-ABJM theory 
with 14 supercharges. But, the Killing spinor equations 
for the six $\CN=2$ supercharges have already determined 
all unknown functions in our ansatz completely. 
Proceeding with the same methods, it is not difficult to show that our solution 
does not admit the other eight `spectator' supercharges.  
We leave the technical proof of this 'no-go' theorem 
and discussion of physical implications for the concluding section.

%\newpage
\section{Details of the Computation \label{detail}}

In this section, we present a detailed analysis 
of Killing spinor equations using the methods 
mentioned in subsection \ref{KSE-meth}, which yielded 
the set of equations summarized in subsection \ref{KSE-sum}.

\subsection{Killing spinor equations}

We want to solve the Killing spinor equation,
\be
\d \psi_m = \nabla_m \e +
\frac{1}{12} \left(\G_m {\bf F} -3 {\bf F}_m \right) \e = 0\,
\;\;\;
\left(\nabla_m\e \equiv \p_m\e + \frac{1}{4} (\w_m)_{ab} \G^{ab} \e \right)\,.
\ee
where we defined, following \cite{oop}, 
\be
{\bf F} \equiv \frac{1}{4!} F_{mnpq} \G^{mnpq} \,,
\;\;\;
{\bf F}_m \equiv \half \left[\G_m , {\bf F} \right] \,.
\ee
Our ansatz for the flux (\ref{flux-ans1}) obviously gives
\be
{\bf F} &=&
e^{-3c_1}\G^{+8}\left(e^{-2c_1}k_1\G^{12}+ e^{4c_1-2h_1}(e^{-2h_3}k_{4,1}\G^{34}+e^{+2h_3}k_{4,2}\G^{56})\right)
\nn \\
&&+e^{h_2}\G^{+9}\left(e^{-2c_1}k_2 \G^{12}+ e^{4c_1-2h_1}(e^{-2h_3}k_{5,1}\G^{34}+e^{+2h_3}k_{5,2}\G^{56})\right)
\nn \\
&&+e^{c_1}\G^{97}\left(e^{-3c_1}k_3 \G^{+8}+ e^{4c_1-2h_1}(e^{-2h_3} k_{6,1}\G^{34}+e^{+2h_3}k_{6,2}\G^{56})\right)
\nn \\
&&+e^{8c_1-4h_1}k_7\G^{3456} \,,
\ee
and (in the orthonormal basis)
\be
{\bf F}_+ &=&
e^{-3c_1}\G^{8}\left(e^{-2c_1}k_1\G^{12}+e^{4c_1-2h_1}(e^{-2h_3}k_{4,1}\G^{34}+e^{+2h_3}k_{4,2}\G^{56})+e^{c_1}k_3\G^{97}\right)
\nn \\
&&+e^{h_2}\G^{9}\left(e^{-2c_1}k_2 \G^{12}+ e^{4c_1-2h_1}(e^{-2h_3}k_{5,1}\G^{34}+e^{+2h_3}k_{5,2}\G^{56})\right) \,,
\nn \\
{\bf F}_- &=& 0 \,,
\nn \\
{\bf F}_1 &=& -\G^{+2}(e^{-5c_1}k_1 \G^{8}+e^{-2c_1+h_2}k_2 \G^{9}) \,,
\nn \\
{\bf F}_2 &=& +\G^{+1}(e^{-5c_1}k_1 \G^{8}+e^{-2c_1+h_2}k_2 \G^{9}) \,,
\nn \\
{\bf F}_3 &=&  +\G^4\left(e^{4c_1-2h_1-2h_3}(e^{-3c_1}k_{4,1}\G^{+8} +e^{h_2}k_{5,1}\G^{+9}+e^{c_1}k_{6,1}\G^{97})+e^{8c_1-4h_1}k_7 \G^{56}\right)\,,
\nn \\
{\bf F}_4 &=&  -\G^3\left(e^{4c_1-2h_1-2h_3}(e^{-3c_1}k_{4,1}\G^{+8} +e^{h_2}k_{5,1}\G^{+9}+e^{c_1}k_{6,1}\G^{97})+e^{8c_1-4h_1}k_7 \G^{56}\right)\,,
\nn \\
{\bf F}_5 &=&  +\G^6\left(e^{4c_1-2h_1+2h_3}(e^{-3c_1}k_{4,2}\G^{+8} +e^{h_2}k_{5,2}\G^{+9}+e^{c_1}k_{6,2}\G^{97})+e^{8c_1-4h_1}k_7 \G^{34}\right)\,,
\nn \\
{\bf F}_6 &=&  -\G^5\left(e^{4c_1-2h_1+2h_3}(e^{-3c_1}k_{4,2}\G^{+8} +e^{h_2}k_{5,2}\G^{+9}+e^{c_1}k_{6,2}\G^{97})+e^{8c_1-4h_1}k_7 \G^{34}\right)\,,
\nn \\
{\bf F}_7 &=& - e^{-2c_1}k_3 \G^{+89} -e^{5c_1-2h_1}\G^9 (e^{-2h_3}k_{6,1}\G^{34}+e^{+2h_3}k_{6,2}\G^{56}) \,,
\nn \\
{\bf F}_8 &=& -e^{-3c_1}\G^+\left(e^{-2c_1}k_1\G^{12}+e^{4c_1-2h_1}(e^{-2h_3}k_{4,1}\G^{34}+e^{+2h_3}k_{4,2}\G^{56})+e^{c_1}k_3 \G^{97}\right) \,,
\nn \\
{\bf F}_9 &=& -e^{h_2}\G^+\left(e^{-2c_1}k_2\G^{12}+e^{4c_1-2h_1}(e^{-2h_3}k_{5,1}\G^{34}+e^{+2h_3}k_{5,2}\G^{56})\right)
\nn \\
&&+e^{c_1}\G^{7}\left(e^{-3c_1}k_3 \G^{+8}+ e^{4c_1-2h_1}(e^{-2h_3}k_{6,1}\G^{34}+e^{+2h_3}k_{6,2}\G^{56})\right) \,.
\ee
We also list the contribution from the spin-connection,
$\w_m \equiv \frac{1}{2} (\w_m)_{ab} \G^{ab}$ (again in the orthonormal basis):
\be
\w_+ &=& -e^{-3c_1}c_2 \G^{8+} +e^{-c_1} \G^{8-} +\thalf e^{h_2}c_2' \G^{9+} -e^{2c_1+h_2}c_1' \G^{9-} -\tfrac{3}{4} e^{c_1}c_3' \G^{97}
\nn \\
&&
+\left\{ \thalf e^{-2c_1}c_2 -\tfrac{3}{4}e^{4c_1-2h_1-2h_3}(1+c_3) \right\} \G^{34}
-\left\{ \thalf e^{-2c_1}c_2 -\tfrac{3}{4}e^{4c_1-2h_1+2h_3}(1-c_3) \right\} \G^{56}
\,,
\nn \\
\w_- &=& \G^{34} - \G^{56} - e^{-c_1} \G^{+8}+e^{2c_1+h_2} c_1' \G^{+9} \,,
\nn \\
\w_1 &=& -e^{-c_1} \G^{18}+  e^{2c_1+h_2} c_1' \G^{19} \,,
\nn \\
\w_2 &=& -e^{-c_1} \G^{28}+  e^{2c_1+h_2} c_1' \G^{29} \,,
\nn \\
\w_3 &=& +\tfrac{3}{4}e^{4c_1-2h_1-2h_3} (1+c_3)\G^{4+} + \thalf e^{5c_1-2h_1+h_2-2h_3} \G^{47} + e^{2c_1+h_2}(-2c_1'+h_1'+h_3')\G^{39} \,,
\nn \\
\w_4 &=& -\tfrac{3}{4}e^{4c_1-2h_1-2h_3} (1+c_3)\G^{3+} - \thalf e^{5c_1-2h_1+h_2-2h_3} \G^{37} + e^{2c_1+h_2}(-2c_1'+h_1'+h_3')\G^{49} \,,
\nn \\
\w_5 &=& -\tfrac{3}{4}e^{4c_1-2h_1+2h_3} (1-c_3)\G^{6+} + \thalf e^{5c_1-2h_1+h_2+2h_3}\G^{67} +e^{2c_1+h_2}(-2c_1'+h_1'-h_3')\G^{59} \,,
\nn \\
\w_6 &=& +\tfrac{3}{4}e^{4c_1-2h_1+2h_3} (1-c_3)\G^{5+} - \thalf e^{5c_1-2h_1+h_2+2h_3} \G^{57} +e^{2c_1+h_2}(-2c_1'+h_1'-h_3')\G^{69} \,,
\nn \\
\w_7 &=& \thalf\left\{ 3e^{-c_1-h_2}(1-c_3) -e^{5c_1-2h_1+h_2-2h_3}\right\} \G^{34}
\nn \\
&&+\thalf\left\{ 3e^{-c_1-h_2}(1+c_3) -e^{5c_1-2h_1+h_2+2h_3}\right\} \G^{56}
\nn \\
&&+\tfrac{3}{4} e^{c_1}c_3' \G^{+9} -e^{2c_1+h_2}(c_1'+h_2') \G^{97} \,,
\nn \\
\w_8 &=& - e^{-c_1} (1-\G^{-}\G^{+}) +e^{2c_1+h_2}c_1' \G^{89} \,,
\nn \\
\w_9 &=& e^{2c_1+h_2} c_1' (1-\G^-\G^+) +\tfrac{3}{4}e^{c_1} c_3' \G^{7+} \,.
\ee

%\newpage

\subsection{Spinorial Lie derivatives}

The Lie derivative of a spinor with respect to a Killing vector is defined by
\be
\Lie_V \e = V^m \nabla_m \e + \frac{1}{4} \left( \nabla_a V_b \right) \G^{ab} \e
\,.
\ee
For the Killing vectors under consideration, the Lie derivatives are given by
\be
&&\Lie_{H} \e = -\p_t \e \,,
\nn \\
&&\Lie_{D} \e = (-2t\p_t -x^i \p_i -r \p_r) \e \,,
\nn \\
&&\Lie_{C} \e = \left[-t^2 \p_t -t(x^i\p_i +r\p_r) - \thalf e^{-c_1} r\G^+(x_i\G^i+r\G^8)
\right] \e
\nn \\
&&\qquad
\;\;\; +\thalf(\vec{x}^2+r^2)\left[ \p_v + \thalf(\G^{34}-\G^{56})\right]\e \,,
\nn \\
&&\Lie_{P_i} \e = \p_i \e \,,
\nn \\
&&\Lie_{G_i} \e = \left[t \p_i + \thalf e^{-c_1}r \G^{+i}\right]\e -x^i \left[\p_v + \thalf (\G^{34}-\G^{56}) \right] \e \,,
\nn \\
&&\Lie_{J} \e = \left[ x_1 \p_2 - x_2\p_1 + \thalf \G^{12} \right] \e \,,
\nn \\
&&\Lie_{M} \e = \left[\p_v +\thalf(\G^{34}-\G^{56}) \right]\e \,,
\nn \\
&&\Lie_{R} \e = \left[\p_w +\thalf(\G^{34}+\G^{56}) \right] \e \,,
\nn \\
&&\Lie_{V_A} \e = V_A \e \,,
\nn \\
&&\Lie_{V'_A} \e = V'_A \e \,.
\label{spinor-Lie}
\ee
We expect $\Lie_{M}\e=0$ for all supercharges, which simplifies
$\Lie_K \e$ and $\Lie_{G_i}\e$ somewhat. 
Here, $V_A$ and $V'_A$ are Killing vectors for the $SU(2)_1 \times 
SU(2)_2$ symmetry (see appendix A). 

The dependence of Killing spinors on each coordinate
is fixed by the Lie derivatives to a large extent.
For dynamical supercharges $Q$, we find
\be
\Lie_H \e_Q = \Lie_{P_i} \e_Q  = \Lie_{V_A} \e_Q  = \Lie_{V'_A} \e_Q  = 0 \,,
\;\; \Lie_D \e_Q  = \e_Q  \;\;\; \imp \;\;\;
\e_Q  = \frac{e^{c_1}}{r} \eta(y) \,,
\label{Q-Lie}
\ee
The fact that $Q$ is singlet under $SU(2)_1\times SU(2)_2$ 
implies that $Q$ is independent of all `three-sphere' coordinates 
($v$, $w$, $\th_i$, $\phi_i$). Then, by $\Lie_M \e =0$, we find 
$\G^{3456}\eta = -\eta$.  
Next, we can use $[G,\bar{Q}]=q$ to get
\be
\e_q = \G^+\left(\frac{\G^1+i\G^2}{2} \right) \eta^c \,,
\label{q-Lie}
\ee
where $\eta^c$ denotes the charge conjugation of $\eta$.
Note that $\G^+\e_q=0$ holds automatically.
Similarly, we can use $[K,Q]=S$ to get
\be
\e_S = \left[\frac{t}{r} e^{c_1} - \half  \G^+(x_i\G^i+r\G^8)\right]\eta \,.
\label{S-Lie}
\ee
All six supercharges $(Q,S,q)$ of the $\CN=2$ subalgebra
are mapped to each other by the bosonic generators.
As a consequence, they all share the same $\eta(y)$
and are independent of ($v$, $w$, $\th_i$, $\phi_i$) coordinates.

\subsection{Kinematical supercharges and null Killing spinor}

\paragraph{G-structure}

In our problem, we have a pair of Killing spinors corresponding to a null Killing vector; recall from (\ref{n2sch3}) that $\{\bar{q}, q\} = 2M$.
The fully general analysis of the geometry with
a single (real) null Killing spinor was done in \cite{gaunt2}.
To import their language, we focus on the real combination
$\e=\thalf(\e_q + \e_{\bar{q}})$ for the moment.

The algebraic relations of \cite{gaunt2} on a null Killing spinor
can be summarized as follows. They take the orthonormal frame
\be
ds^2 = 2 e^+e^- + e^ie^i + e^9e^9  \,,
\label{G-frame}
\ee
with $i=1,\cdots,8$ and
\be
K = e^+ \,.
\ee
They further choose the Killing spinor to satisfy
\be
\label{null-proj}
\G_{1234} \e = \G_{3456} \e = \G_{5678} \e = \G_{1357} \e = -\e \,,
\;\;\;
\G^+ \e = 0\,.
\ee
These conditions automatically implies $\G^9\e=\e$.
Next, they show that this spinor defines a Spin$(7)$ structure
within the $\IR^8$ subspace of the tangent bundle.
In particular, they find
\be
\Om = e^+ \wedge e^9 \,, \;\;\;
\S = e^+ \wedge \Phi \,,
\label{spin7-cano}
\ee
where $\Phi$ is the invariant four form defining
the embedding of Spin(7) into Spin(8),
\be
-\Phi &=& e^{1234}+e^{1256}+e^{1278}+e^{3456}+ e^{3478}+e^{5678}
\nn \\
&& +e^{1357}-e^{1368}-e^{1458}-e^{1467}-e^{2358}-e^{2367}-e^{2457}+e^{2468} \,.
\ee

Our choice of
the orthonormal frame (\ref{ortho}) is such that $e^+$ is the dual one-form of $M$ in agreement with \cite{gaunt2}. We also showed already that
parts of the conditions (\ref{null-proj}), namely, $\G^{3456}\e=-\e$ and
$\G^+\e=0$ hold for the kinematical supercharges.
On the other hand, it is not clear whether
the $(8+1)$ splitting in the canonical G-structure frame (\ref{G-frame})
agrees with our original choice of the frame (\ref{ortho}).
In fact, we will see that the two frames are
related to each other by a $y$-dependent rotation on the $(89)$-plane.

%\subsubsection{Solving KSE with hints from G-structure}

\paragraph{Killing spinor equations}

We showed earlier that the kinematical supercharges satisfy
\be
\G^+ \e = 0 \,,
\;\;\;
\G^{3456} \e = -\e \,,
\;\;\;
\p_m \e = 0 \;\;\; (\mbox{except for }m=y)\,.
\ee
These conditions simplify the KSE drastically:
\be
\d\psi_{1,2,8} \; &\imp& \; \Pi_1 \e \equiv
\bigg[ -e^{-c_1} \G^8 + e^{2c_1+h_2} c_1' \G^9
\nn \\
&& \hskip 1.6cm
+ \tfrac{1}{6}e^{5c_1-2h_1} \G^{97}(e^{-2h_3}k_{6,1}\G^{34}+e^{2h_3}k_{6,2} \G^{56})
%\nn \\
%&& \; \hskip 8.0cm
-\tfrac{1}{6}e^{8c_1-4h_1}k_7  \bigg] \e = 0 \,,
\nn \\
\G^3\d\psi_3 + \G^5 \d\psi_5  \; &\imp& \; \Pi_2 \e \equiv
\bigg[ 2e^{2c_1+h_2}(-2c_1'+h_1')\G^9 + \thalf e^{5c_1-2h_1+h_2} \G^7 (e^{-2h_3}\G^{34} + e^{+2h_3}\G^{56})
\nn \\
&& \;  \hskip 1.5cm -\tfrac{1}{6}e^{5c_1-2h_1} \G^{97}(e^{-2h_3}k_{6,1}\G^{34}+e^{2h_3}k_{6,2} \G^{56})
+\tfrac{2}{3}e^{8c_1-4h_1}k_7  \bigg] \e =0 \,,
\nn \\
\G^3\d\psi_3 - \G^5 \d\psi_5  \; &\imp& \; \Pi_3 \e \equiv
\bigg[ 2e^{2c_1+h_2}h_3'\G^9 +\thalf e^{5c_1-2h_1+h_2} \G^7 (e^{-2h_3}\G^{34} - e^{2h_3}\G^{56})
\nn \\
&& \; \hskip 3.5cm -\tfrac{1}{2}e^{5c_1-2h_1} \G^{97}(e^{-2h_3}k_{6,1}\G^{34}-e^{2h_3}k_{6,2} \G^{56})  \bigg] \e =0 \,,
\nn \\
\d\psi_7 \; &\imp& \; \Pi_4 \e \equiv
\bigg[ e^{2c_1+h_2}(c_1'+h_2')\G^9 +\tfrac{3}{2}e^{-c_1-h_2}\G^7(\G^{34}+\G^{56})
\nn \\
&& \hskip 1.6cm - \thalf e^{5c_1-2h_1+h_2} \G^7 (e^{-2h_3}\G^{34} + e^{2h_3}\G^{56})
\nn \\
&& \hskip 1.6cm -\tfrac{1}{3} e^{5c_1-2h_1} \G^{97}(e^{-2h_3}k_{6,1}\G^{34}+e^{2h_3}k_{6,2} \G^{56})
-\tfrac{1}{6}e^{8c_1-4h_1}k_7  \bigg] \e = 0 \,,
\nn \\
\d\psi_{9} \; &\imp& \; \Pi_5 \e \equiv
\bigg[ 2e^{2c_1+h_2} \G^9 \p_y +  e^{2c_1+h_2} c_1' \G^9
\nn \\
&& \hskip 1.6cm
- \tfrac{1}{3} e^{5c_1-2h_1} \G^{97}(e^{-2h_3}k_{6,1}\G^{34}+e^{2h_3}k_{6,2} \G^{56})
%\nn \\
%&& \; \hskip 8.0cm
-\tfrac{1}{6}e^{8c_1-4h_1}k_7  \bigg] \e = 0 \,,
\nn \\
\d\psi_+  \; &\imp& \; \Pi_6 \e \equiv \bigg[ \thalf \mathbf{F}_++ \tfrac{3}{4} e^{c_1}c_3'\G^{97}
\nn \\
&& \hskip 1.5cm +\tfrac{3}{4}e^{4c_1-2h_1} ( e^{-2h_3}(1+c_3)\G^{34} - e^{2h_3} (1-c_3)\G^{56} ) \bigg] \e = 0 \,.
\label{KSE-kin}
\ee
The last equation, $\d\psi_- = 0$, is equivalent to $\Lie_M \e = 0$.

%\paragraph{Projectors?}
%We can imitate the first step in the Bena-Warner analysis,
%\be
%4(\G^1 \d\psi_1 + \G^3 \d\psi_3 + \G^5 \d\psi_5 )_{F-{\rm dep}} =
%{\bf F} - \G^1{\bf F}_1 - \G^3{\bf F}_3- \G^5{\bf F}_5
%= k_3 \G^{+789} - k_7 \G^{3456} \,.
%\ee
%As a first attempt, it may be useful to assume $k_3=k_7=0$ and
%try to see whether we can find any solution or contradiction.

\paragraph{Computation of $\Om$ and determination of $(e^{9'})$}
We can use the projection conditions to compute various components of
$\Om$.
Assuming for now that $k_7\neq 0$, we find
\be
\bar{\e} \G_{+78} (\Pi_1+\Pi_2)\e = 0 \;&\imp&\; \Om_{+7} = 0 \,,
\nn \\
\bar{\e} \G_{+9} \Pi_1 \e = 0 \;&\imp&\; \Om_{+9} = \frac{6c_1'}{k_7} e^{-6c_1+4h_1+h_2} \,,
\nn \\
\bar{\e} \G_{+} (3\Pi_1+\Pi_2+\Pi_4) \e = 0 \;&\imp&\; \Om_{+8} = \frac{2c_1'}{k_7} (2h_1'+h_2') e^{-3c_1+4h_1+2h_2} \,.
\label{Om-det}
\ee
It is also easy to show that $\Om_{+i}=0$
$(i=1,\cdots,6)$ in a similar way.
Thus, we arrive at the canonical form of $\Om = e^+ \wedge e^{9'}$
upon the following rotation among vielbein:
\be
\begin{pmatrix}
e^{8'} \\ e^{9'}
\end{pmatrix}
=
\begin{pmatrix}
\cos\zeta & \sin\zeta \\ -\sin\zeta & \cos\zeta
\end{pmatrix}
\begin{pmatrix}
e^{8} \\ e^{9}
\end{pmatrix}
\,,
\label{frame-rot}
\ee
where
\be
\cos\zeta = \frac{6c_1'}{k_7} e^{-6c_1+4h_1+h_2}  \,,
\;\;\;
\sin\zeta = -\frac{2c_1'}{k_7} (2h_1'+h_2') e^{-3c_1+4h_1+2h_2} \,.
\label{rot-angle}
\ee
{}From $\cos^2\zeta +\sin^2\zeta = 1$, we find a non-trivial condition
among the unknown functions,
\be
4(c_1')^2 e^{-12c_1+8h_1+2h_2} \left[ 9 + (2h_1'+h_2')^2 e^{6c_1+2h_2} \right] = k_7^2 \,.
\ee
Another non-trivial relation follows from
\be
\bar{\e} \G_{+} (2\Pi_2-\Pi_4) \e = 0 \;&\imp&\; k_7^2 =4c_1'(9c_1'-4h_1'+h_2')e^{-12c_1+8h_1+2h_2} \,,
\label{om-last}
\ee
where we used the form of $\Om_{+9}$ in (\ref{Om-det}).
We can eliminate $k_7$ from the two equations above to
obtain a relation among metric components only
\be
4h_1'-h_2' = -c_1'(2h_1'+h_2')^2 e^{6c_1+2h_2} \,.
\label{combo1}
\ee

\paragraph{Computation of $\S$ and confirmation of $(e^{9'})$}
We can also use the projection conditions to compute components of
$\S$. For instance, we find
\be
\bar{\e} \G_{+9} (\Pi_2+\Pi_3)\e = 0 \;&\imp&\;
\S_{+7934} = 2e^{-3c_1+2h_1+2h_3} (h_1'+h_3') \,,
\nn \\
\bar{\e} \G_{+9} (\Pi_2-\Pi_3) \e = 0 \;&\imp&\;
\S_{+7956} = 2e^{-3c_1+2h_1-2h_3} (h_1'-h_3') \,,
\label{Sigma1}
\ee
It follows from $\S_{+ij34} = \S_{+ij56}$ that
\be
h_1'\sinh(2h_3) + h_3' \cosh(2h_3) =0 \,.
\label{a4x}
\ee
%This condition is trivial if $h_3=0$ identically.
%Otherwise, we can integrate it to find
%\be
%|\sinh(2h_3)| = \k_1 \, e^{-2h_1} \,,
%\label{f2-sol}
%\ee
%where $\k_1$ is an integration constant.
Next, we can determine $\{k_{6,a}\}$. Note that
\be
\bar{\e} \G_{+} (3\Pi_2+\Pi_3)\e = 0 \;&\imp&\;
k_{6,1} \S_{+7934} = -\frac{4c_1'}{k_7}(h_1'+2h_2'+3h_3') e^{-9c_1+6h_1+2h_2+2h_3} \,,
\nn \\
\bar{\e} \G_{+} (3\Pi_2-\Pi_3) \e = 0 \;&\imp&\;
k_{6,2} \S_{+7956} = -\frac{4c_1'}{k_7}(h_1'+2h_2'-3h_3') e^{-9c_1+6h_1+2h_2-2h_3} \,.
\ee
Then, using (\ref{Sigma1}), we find
\be
k_{6,1} &=& -\frac{2c_1'(h_1'+2h_2'+3h_3')}{k_7(h_1'+h_3')}e^{-6c_1+4h_1+2h_2} \,,
\nn \\
k_{6,2} &=& -\frac{2c_1'(h_1'+2h_2'-3h_3')}{k_7(h_1'-h_3')}e^{-6c_1+4h_1+2h_2} \,.
\label{k-six}
\ee
{}From $\S = e^+\wedge \Phi$ with the rotation taken into account, we deduce
\be
&&\S_{+7934} = -\sin\zeta
\nn \\
\imp \; && \;
2e^{2h_3} (h_1'+h_3')
= \frac{2c_1'}{k_7} (2h_1'+h_2') e^{2h_1+2h_2} \,.
\label{S-sin}
\ee

\paragraph{Constancy of spinor and further relations}

Recall that the canonical G-structure frame (where $\Om = e^+\wedge e^{9'}$ holds) is related to our original frame (\ref{ortho}) by the rotation (\ref{frame-rot}). {}Since the projection conditions are mapped
to each other by
\be
(-\sin\zeta \G^8 + \cos\zeta \G^9)\e = \e \;\;\; \Longleftrightarrow \;\;\;
\G^9 \e' = \e'  \,,
\label{pro-map}
\ee
the Killing spinors in the two frames should be related by
\be
\e' = \exp\left[(\zeta/2) \G^{89} \right] \e
= \left[ \cos(\zeta/2)+\G^{89} \sin(\zeta/2) \right] \e \,.
\label{ep-rot}
\ee
We can now take advantage of another important result of \cite{gaunt2}
that $\e'$ is a {\em constant} spinor.
Plugging (\ref{ep-rot}) this into the $\d\psi_9=0$ condition in (\ref{KSE-kin})
and using $d\e'=0$, we obtain
\be
\Pi_5 \e &=& \bigg[ \e^{2c_1+h_2} \zeta' \G^8 + e^{2c_1+h_2} c_1' \G^9
- \tfrac{1}{3}e^{5c_1-2h_1} \G^{97}(e^{-2h_3}k_{6,1}\G^{34}+e^{2h_3}k_{6,2} \G^{56})
\nn \\
&& \; \hskip 8.0cm-\tfrac{1}{6}e^{8c_1-4h_1}k_7 \bigg] \e = 0 \,,
\ee
which further implies
\be
&&\Pi_1' \e \equiv (\Pi_1 -\Pi_5)\e
\nn \\
&& \hskip 0.8cm
= \left[ (-\zeta'e^{2c_1+h_2}-e^{-c_1})\G^8 +\thalf e^{5c_1-2h_1}\G^{97}(e^{-2h_3}k_{6,1} \G^{34} + e^{2h_3}k_{6,2} \G^{56} ) \right] \e = 0 \,,
\nn \\
&&\Pi_4' \e \equiv (\Pi_4 -\Pi_5)\e = \bigg[-\zeta' e^{2c_1+h_2} \G^8 +h_2'e^{2c_1+h_2} \G^9 +\tfrac{3}{2}e^{-c_1-h_2}\G^7(\G^{34}+\G^{56})
\nn \\
&&\hskip 6cm
- \thalf e^{5c_1-2h_1+h_2} \G^7 (e^{-2h_3}\G^{34} + e^{2h_3}\G^{56})\bigg] \e =0 \,,
\nn \\
&&\Pi_0 \e \equiv -\frac{e^{-8c_1+4h_1}}{k_7} (2\Pi_1 + \Pi_5) \e
\nn \\
&& \; \hskip 0.7cm = \half \left[ 1 + \frac{2}{k_7}e^{-6c_1+4h_1}\left( (-\zeta'e^{h_2}+2e^{-3c_1})\G^8 - 3c_1'e^{h_2} \G^9 \right)\right] \e = 0 \,.
\ee
Comparing $\Pi_0$ with (\ref{rot-angle}) and (\ref{pro-map}), we make
the identification,
\be
\sin\zeta = \frac{2}{k_7}(-\zeta'e^{h_2} + 2e^{-3c_1})e^{-6c_1+4h_1} \,.
\label{pro-iden}
\ee
Next, we have
\be
\bar{\e} \G_+ \Pi'_4 \e = 0 \; \imp \; \zeta' \sin\zeta = -h_2'\cos \zeta
\,.
\label{xi-eq}
\ee
Integrating it and using (\ref{rot-angle}) again, we find
\be
\cos\zeta = \k \,e^{h_2} = \frac{6c_1'}{k_7} e^{-6c_1+4h_1+h_2}
\;\; \imp \;\;
k_7 = \frac{6c_1'}{\k} e^{-6c_1+4h_1} \,,
\label{k7-eq}
\ee
where $\k$ is an integration constant.
Combining it with (\ref{om-last}), we also find
\be
9c_1' e^{-2h_2} = \k (9c_1'-4h_1'+h_2')  \,.
\label{more1}
\ee
As a consistency check, we combine (\ref{rot-angle}), (\ref{pro-iden}) and (\ref{xi-eq}) to find
\be
4h_1'-h_2'=-c_1'(2h_1'+h_2')^2e^{6c_1+2h_2}\,,
\label{combo2}
\ee
which is identical to (\ref{combo1}).

Combining $\Pi_1'$ with (\ref{k-six}), (\ref{pro-iden}) and (\ref{k7-eq}) gives more projection conditions and constraints on unknown functions
\be
\Pi_1' \e = 0 \;&\imp&\; \G^{34789} \e = -\e \,, \quad 2h_1'+h_2' = 6 \k (h_1'+h_3')e^{-6c_1+2h_1-2h_2+2h_3} \,.
\label{pi1-cond}
\ee
and we can easily check $\Pi_{2,3}$ conditions are automatically satisfied with (\ref{pi1-cond}). The remaining projection conditions, $\Pi_4', \Pi_6$, produce the following conditions
\be
\Pi_4' \e = 0 \;&\imp&\; \k = 1 \,,
\label{pi4p}
\\
\Pi_6 \e = 0 \;&\imp&\; +\sin \zeta\left(e^{-2c_1}k_3+3e^{4c_1-2h_1}(c_3 \cosh(2h_3) -\sinh(2h_3))\right)
\nn \\
&& \; \hskip 0.5cm =\tfrac{3}{2}c_3'e^{c_1}+e^{-3c_1}\left(e^{-2c_1}k_1+e^{4c_1-2h_1}(e^{-2h_3}k_{4,1}+e^{2h_3}k_{4,2})\right) \,,
\nn \\
&& \; -\cos \zeta\left(e^{-2c_1}k_3+3e^{4c_1-2h_1}(c_3 \cosh(2h_3) -\sinh(2h_3))\right)
\nn \\
&& \;\hskip 0.5cm = e^{h_2}\left(e^{-2c_1}k_2+e^{4c_1-2h_1}(e^{-2h_3}k_{5,1}+e^{2h_3}k_{5,2})\right) \,.
\label{pi46-cond}
\ee

\paragraph{Summary}
We have found that the null Killing spinor equations
impose several coupled equations for the unknown functions
$\{c_1,h_1,h_2,h_3,k_{6,1},k_{6,2},k_7\}$.
The independent equations can be summarized as follows:
\be
&& 4h_1' -h_2' = -c_1'(2h_1'+h_2')^2 e^{6c_1+2h_2} \,,
\label{a1}
\nn \\
&& 9c_1 ' = (9c_1'-4h_1' +h_2')e^{2h_2} \,,
\label{a2}
\nn \\
&&2h_1'+h_2' = 6(h_1'+h_3') e^{-6c_1 +2h_1-2h_2+2h_3} \,,
\label{a3}
\nn \\
&& h_3' \cosh(2h_3) = -h_1' \sinh(2h_3)
\,,
\label{a4}
\nn \\
&&
k_{6,1} = - \frac{h_1'+2h_2'+3h_3'}{3(h_1'+h_3')} e^{2h_2} \,,
\nn \label{c4a}
\\
&&
k_{6,2} = - \frac{h_1'+2h_2'-3h_3'}{3(h_1'-h_3')} e^{2h_2} \,,
\nn \label{c4b}
\\
&&
k_7 = 6c_1' e^{-6c_1+4h_1} \,.
\nn \label{c5}
\ee
The first four equations were obtained in (\ref{combo1}), (\ref{more1}),
(\ref{pi1-cond}) and (\ref{a4x}), respectively;
recall also (\ref{pi4p}). They give Block A in section 3. 
The auxiliary equations in Block A concerning $\cos\z$ and $\sin\z$ come from 
combinations of (\ref{k7-eq}), (\ref{pi4p}), (\ref{rot-angle}) and (\ref{pro-iden}). 
The equations for $(k_{6,1}, k_{6,2}, k_7)$ were obtained in (\ref{k-six})
and (\ref{k7-eq}). They give the last three entries of Block C in section 3.

%Let us recall that we began our analysis with
%\be
%\G^{+} \e = 0 \,, \;\;\;\G^{3456} \e = -\e \,.
%\ee
%During the analysis, we have arrived at
%\be
%(-\sin\zeta \G^8 + \cos\zeta \G^9)\e = \e \;\;\; \Longleftrightarrow \;\;\;
%\G^9 \e' = \e'  \,.
%\ee
%Finally, our orientation convention conforms to that of \cite{gaunt2}; see (\ref{S-sin}),
%\be
%\G^{3478} \e' = -\e' \,.
%\ee

%\newpage
\subsection{Dynamical supercharges and time-like Killing spinor}

\paragraph{G-structure}
The commutation relation $\{\bar{Q} , Q\}=H$ implies that 
an $\CN=2$ super-Schr\"odinger geometry should admit a time-like Killing spinor. 
The general study of geometries admitting a single time-like Killing spinor 
has been done in \cite{gaunt1}. The metric takes the form 
\be
ds^2 = -\D^2(dt+\w)^2 + \D^{-1} g_{mn} dx^m dx^n \,.
\ee
The base manifold $\CB$ with metric $g_{mn}$ is orthogonal to time direction and has $SU(5)$ structure instead of Spin(7). The $SU(5)$ structure is given by a pair of spinors $(\e_{\bar{Q}},\e_Q)$.
\be
\e_d &\equiv& \tfrac{1}{\sqrt{2}}(\e_{\bar{Q}}+\e_Q)
\nn \\
K&=& \bar\epsilon_d \G_a\epsilon_d \, e^a = \D^2(dt+\omega)
\nn \\
\Omega &=& \bar\epsilon_d \G_{ab}\epsilon_d \, e^{ab}
\ee
Here, $K$ is the dual one-form of $H$ and $\Omega$ is the K\"ahler form of the base manifold B. We can always decompose
$\epsilon$ by the eigenvalue of $\G^{+-}$ and find the relation with the kinematical supercharges
\be
\e_d &=&\frac{e^{c_1}}{r}\left( \G^+ \eta_1 + \eta_2 \right) \,,
\nn \\
\e_k &\equiv& \thalf (\e_{q}+\e_{\bar{q}}) = \tfrac{1}{\sqrt{2}} \G^{+1} \eta_2 \,, 
\label{dy-decomp}
\ee
where
\be
&&\G^-\eta_i = 0 \,, \;\;\;\;
\G^{3456}\eta_i = -\eta_i \,, \quad\qquad (i=1,2)
\nn \\
&&\bar\eta_2\G_-\eta_2 = -1 \,,\;\;\;\;
\bar\eta_1\G_-\eta_1 = -\tfrac{1}{4} c_2 e^{-2c_1} \,.
\ee
The last two relations are derived from the relation of the spinor bilinear $K$. 
The $\eta_i$'s are orthogonal to each other and have zero-norm $\bar\eta_i \eta_j = 0$.

%\subsubsection{KSE and G-structure}
\paragraph{Killing spinor equations}
{}From the results of the previous subsection, 
we already have some information about the dynamical supercharge
\be
\G^{3456}\e_d = -\e_d \,, \quad r\partial_r \e_d = -\e_d \,, \quad \partial_m\e_d = 0 \quad (\mbox{except for }m = y, r)
\ee
The Killing spinor equations for the component spinors $\eta_i$ are given by
\be
&&\bullet \; \delta \psi_{1,2}
\nn \\
&&	\Rightarrow \left[e^{-c_1}\Gamma^8-e^{2c_1+h_2}c_1'\Gamma^9+\tfrac{1}{6}e^{5c_1-2h_1}\Gamma^{97}(e^{-2h_3}k_{6,1}\Gamma^{34}+e^{2h_2}k_{6,2}\Gamma^{56})-e^{8c_1-4h_1}\tfrac{1}{6}k_7\right]\eta_1
\nn \\
&& \, \hskip 10cm +\tfrac{1}{6}\left[{\bf F}_+-\tfrac{3}{2}\G^{-1}{\bf F}_1\right] \eta_2 = 0 \;,
\nn \\
&&\bullet \; \Gamma^3\delta\psi_3 + \Gamma^5\delta\psi_5
\nn \\
&&\Rightarrow \bigg[-2(-2c_1'+h_1')e^{2c_1+h_2}\Gamma^9 - \thalf e^{5c_1-2h_1+h_2}\Gamma^7(e^{-2h_3}\Gamma^{34}+e^{2h_3}\Gamma^{56})
\nn \\
&& \, \hskip 4.0cm -\tfrac{1}{6}e^{5c_1-2h_1}\Gamma^{97}(e^{-2h_3}k_{6,1}\Gamma^{34}+e^{2h_3}k_{6,2}\Gamma^{56})+\tfrac{2}{3}e^{8c_1-4h_1}k_7\bigg] \eta_1
\nn \\
&&\quad+\left[\tfrac{3}{4}e^{4c_1-2h_1}(e^{-2h_3}(1+c_3)\Gamma^{34}-e^{2h_3}(1-c_3)\Gamma^{56}) \right.
\nn \\
&& \hskip5cm \left. +\tfrac{1}{6}\left(2{\bf F}_+-\tfrac{3}{2}(\G^{-3}{\bf F}_3+\G^{-5}{\bf F}_5\right)\right]\eta_2 = 0 \;,
\nn \\
&&\bullet \; \Gamma^3\delta\psi_3 - \Gamma^5\delta\psi_5 \nn \\
&&\Rightarrow \left[-2h_3'\,e^{2c_1+h_2}\Gamma^9 - \thalf e^{5c_1-2h_1+h_2}\Gamma^7(e^{-2h_3}\Gamma^{34}-e^{2h_3}\Gamma^{56}) \right.
\nn \\
&&\hskip 3cm \left. -\thalf e^{5c
_1-2h_1}\Gamma^{97}(e^{-2h_3}k_{6,1}\Gamma^{34}-e^{2h_3}k_{6,2}\Gamma^{56})\right] \eta_1 \nn \\
&&\hskip 1cm +\left[\tfrac{3}{4}e^{4c_1-2h_1}(e^{-2h_3}(1+c_3)\Gamma^{34}+e^{2h_3}(1-c_3)\Gamma^{56}) -\tfrac{1}{4}(\G^{-3}{\bf F}_3-\G^{-5}{\bf F}_5)\right]\eta_2 = 0\;,
\nn \\
&&\bullet \; \delta \psi_7
\nn \\
&&\Rightarrow \left[-e^{2c_1+h_2}(c_1'+h_2')\Gamma^9-\tfrac{3}{2}e^{-c_1-h_2}\Gamma^7(\Gamma^{34}+\Gamma^{56})+\thalf e^{5c_1-2h_1+h_2}\Gamma^7(e^{-2h_3}\Gamma^{34}+e^{2h_3}\Gamma^{56}) \right.
\nn \\
&&\, \hskip 4.0cm \left. -\tfrac{1}{3}e^{5c_1-2h_1}\Gamma^{97}(e^{-2h_3}k_{6,1}\Gamma^{34}+e^{2h_3}k_{6,2}\Gamma^{56})-\tfrac{1}{6}e^{8c_1-4h_1}k_7\right]\eta_1
\nn \\
&& \, \hskip 6.0cm+\left[-\tfrac{3}{4} e^{c_1}c_3'\Gamma^{79}+\tfrac{1}{6}\left({\bf F}_+ -\tfrac{3}{2}\G^{-7}{\bf F}_7\right) \right]\eta_2 = 0 \;,
\nn \\
&&\bullet \; \delta \psi_8
\nn \\
&&\Rightarrow \left[3e^{-c_1}\Gamma^8 - e^{2c_1+h_2}c_1'\Gamma^9 +\tfrac{1}{6}e^{5c_1-2h_1}\Gamma^{97}(e^{-2h_3}k_{6,1}\Gamma^{34}+e^{2h_3}k_{6,2}\Gamma^{56})-\tfrac{1}{6}e^{8c_1-4h_1}k_7\right]\eta_1
\nn \\
&& \, \hskip 8.7cm +\tfrac{1}{6}\left[{\bf F}_+ -\tfrac{3}{2}\G^{-8}{\bf F}_8\right]\eta_2 = 0 \;,
\nn \\
&&\bullet \; \delta \psi_9
\nn \\
&&\Rightarrow \left[-2e^{2c_1+h_2}\Gamma^9\partial_y -3e^{2c_1+h_2}c_1'\Gamma^9
\right.
\nn \\
&& \hskip 2cm \left.-\tfrac{1}{3}e^{5c_1-2h_1}\Gamma^{97}(e^{-2h_3}k_{6,1}\Gamma^{34}+e^{2h_3}k_{6,2}\Gamma^{56})-\tfrac{1}{6}e^{8c_1-4h_1}k_7\right]\eta_1
\nn \\
&& \, \hskip 6.1cm +\left[-\tfrac{3}{4}e^{c_1}c_3'\Gamma^{79}+\tfrac{1}{6}({\bf F}_+ -\tfrac{3}{2}\G^{-9}{\bf F}_9)\right]\eta_2 = 0 \;,
\nn \\
&&\bullet \; \delta \psi_+
\nn \\
&&\Rightarrow \left[\tfrac{3}{4}e^{c_1}c_3'\Gamma^{79}-\tfrac{3}{4}e^{4c_1-2h_1}(e^{-2h_3}(1+c_3)\Gamma^{34}-e^{2h_3}(1-c_3)\Gamma^{56})+\tfrac{1}{2}{\bf F}_+\right]\eta_1 \nn \\
	&&\, \hskip 7.2cm +\left[c_2e^{-3c_1}\Gamma^8-\thalf c_2'e^{h_2}\Gamma^9\right]\eta_2 = 0 \,.
\label{dy-kse}
\ee
The equations for $\delta\psi_1$, $\delta\psi_8$ give a relation between $\eta_1$ and $\eta_2$,
\be
\eta_1 = \tfrac{1}{4}\left(e^{2c_1-2h_1}(e^{-2h_3}k_{4,1}+e^{2h_3}k_{4,2})\G^{34}-e^{-c_1}k_3\G^{79}-e^{-c_1+h_2}k_2\G^{1289}\right)\eta_2 \,.
\ee

\paragraph{Sufficiency of Killing spinor equations}
In general, Killing spinor equations do not restrict every single component of the metric and flux. To determine all components, we must supplement 
the Killing spinor equations with some components of the equation of motion. 
However, the situation is better for our problem. 
Note that we have the time-like Killing vector $H$ 
as well as the null Killing vector $M$. In addition, our flux does not have 
components along the $e^{-}$ direction. These facts together imply that 
in our case, the Killing spinor equations are sufficient to determine 
all components of the metric and flux \cite{gaunt1}. 

\paragraph{Computation of $\Omega$ and further constraints}
We wrote the dynamical Killing spinor in terms of $\eta_2$ which inherit the properties of the kinematical Killing spinor $\e_k$. The projection conditions on $\eta_2$ are summarized by
\be
&&\G_-\eta_2 = 0 \,, \quad \G^{3456}\eta_2 = -\eta_2 \,, \quad \G^{34789}\eta_2 = -\eta_2 \,,
\nn \\
&&(-\sin \zeta \G^8 + \cos \zeta \G^9)\eta_2 = \eta_2 \, .
\label{dy-proj}
\ee
Using these and the relations (\ref{dy-decomp}), 
we can derive the explicit form of the K\"ahler form $\Omega$
\be
\Omega &=& \frac{1}{2r^2}\left[-(c_2 e^+ + 2e^{2c_1} \, e^-)e^{9'}\right.
\nn \\
&&\quad \quad +\left(e^{4c_1-2h_1}(e^{-2h_3}k_{4,1}+e^{2h_3}k_{4,2})-e^{c_1}k_3\sin \zeta \right)(-e^{12}+e^{34}+e^{56})
\nn \\
&& \quad \quad +\left(e^{4c_1-2h_1}(e^{-2h_3}k_{4,1}+e^{2h_3}k_{4,2})\cos \zeta + e^{c_1+h_2}k_2\sin \zeta \right)e^{78}
\nn \\
&& \left. \quad \quad +\left(e^{4c_1-2h_1}(e^{-2h_3}k_{4,1}+e^{2h_3}k_{4,2})\sin \zeta - e^{c_1+h_2}k_2\cos \zeta - e^{c_1}k_3\right)e^{79}\right] \,.
\ee
We used the condition $i_K\Omega=0$ to derive $\Omega_{+a}$ components. 
The G-structure equations give further relations
\be
&& \Omega_a^{\;\;c}\Omega_c^{\;\;b} = -K_aK^b+\delta_a^{\;b}K^2
\nn \\
&& \imp \;\;\; k_2=-k_3 \,, \quad c_2 = \tfrac{1}{4}\left(e^{3c_1-2h_1}(e^{-2h_3}k_{4,1}+e^{2h_3}k_{4,2})- k_3\sin \zeta\right)^2 \,.
\ee
{}From $d\Omega =i_K F$, 
\be
k_1&=&-2\left[e^{6c_1-2h_1}(e^{-2h_3}k_{4,1}+e^{2h_3}k_{4,2})-e^{3c_1}k_3\sin\zeta \right] \,,
\nn \\
k_3&=& -\tfrac{3}{2} c_3'e^{3c_1}\sin\zeta+\thalf e^{-3c_1}k_1\sin\zeta+ \tfrac{1}{4}\partial_y(e^{2h_2}k_1) \,,
\nn \\
k_{4,1} &=& \tfrac{3}{2}\left[-(c_3+1)e^{3c_1}\sin\zeta-\tfrac{1}{6}k_1(2e^{-6c_1+2h_1+2h_3}-e^{2h_2})\right] \,,
\nn \\
k_{4,2} &=& \tfrac{3}{2}\left[-(c_3-1)e^{3c_1}\sin\zeta-\tfrac{1}{6}k_1(2e^{-6c_1+2h_1-2h_3}-e^{2h_2})\right] \,,
\nn \\
k_{5,1} &=& \tfrac{1}{4}\left[6(c_3+1)+e^{-3c_1}\sin\zeta+\partial_y(e^{-6c_1+2h_1+2h_3}k_1)\right] \,,
\nn \\
k_{5,1} &=& \tfrac{1}{4}\left[6(c_3-1)+e^{-3c_1}\sin\zeta+\partial_y(e^{-6c_1+2h_1-2h_3}k_1)\right] \,.
\ee
We can use the Bianchi identity of $k_1'=-4k_2$ and (\ref{f2-sol}), (\ref{k7-eq}) to simplify the equations for $k_3, k_{5,i}$ such that
\be
&&3c_3' = 2\left(k_1e^{-6c_1}\frac{h_1'-h_2'}{2h_1'+h_2'}+k_3e^{-3c_1}\sin \zeta\right) \,,
\nn \\
&&e^{-2h_3}k_{5,1}-e^{2h_3}k_{5,2}=-3(c_3\sinh(2h_3)-\cosh(2h_3)) \,.
\ee
By combining these results, we can 
further reduce the equation (\ref{pi46-cond}) to a simpler form,
\be
&&k_{5,1} = -\tfrac{3}{2}(c_3-1) \,, \qquad
k_{5,2} = -\tfrac{3}{2}(c_3+1) \,,
\nn \\
&&3c_1'+k_1e^{-6c_1} = 6\sin\zeta(c_3 \cosh(2h_3)-\sinh(2h_3))e^{3c_1-2h_1} \,.
\ee
It is straightforward to show that the solutions which satisfy all the equations 
we have found so far will also satisfy the rest of Killing spinor equations.

\paragraph{Summary}
We can find one more relation for $k_1$ and $c_3$ from the super-Schr\"odinger algebra; our solution for the Killing spinor realizes the commutation relation (\ref{n2sch3}) 
if 
\be
k_1 = -\frac{6c_3}{\sin\zeta}e^{3c_1} \,.
\ee
Thus we have found all equations in Block B,
\be
&&k_2=-k_3 \,,
\nn \\
&&k_1 = -\frac{6c_3}{\sin\zeta}e^{3c_1} \,,
\nn \\
&&3c_1'+k_1e^{-6c_1}=6\sin\zeta(c_3 \cosh(2h_3)-\sinh(2h_3))e^{3c_1-2h_1} \,,
\nn \\
&&3c_3' = 2\left(k_1e^{-6c_1}\frac{h_1'-h_2'}{2h_1'+h_2'}+k_3e^{-3c_1}\sin \zeta\right) \,,
\ee
and the first five entries of Block C,
\be
&&c_2 = (\tfrac{1}{4}k_1e^{-3c_1})^2 \,,
\nn \\
&&k_{4,1} = \tfrac{3}{2}\left[-(c_3+1)e^{3c_1}\sin\zeta-\tfrac{1}{6}k_1(2e^{-6c_1+2h_1+2h_3}-e^{2h_2})\right] \,,
\nn \\
&&k_{4,2} = \tfrac{3}{2}\left[-(c_3-1)e^{3c_1}\sin\zeta-\tfrac{1}{6}k_1(2e^{-6c_1+2h_1-2h_3}-e^{2h_2})\right] \,,
\nn \\
&&k_{5,1}=-\tfrac{3}{2}(c_3-1) \,,
\nn \\
&&k_{5,2}=-\tfrac{3}{2}(c_3+1) \,.
\ee
The dynamical Killing spinor $\e_d$ and K\"ahler form $\Omega$ is reduced to
\be
\e_d &=& \frac{e^{c_1}}{r}(\G^+\eta_1+\eta_2) = \frac{e^{c_1}}{r}(-\tfrac{1}{8}e^{-4c_1}k_1\G^{+34}+1)\eta_2 \,,
\nn \\
\Omega &=& -\frac{1}{2r^2}(c_2 e^+ +2e^{2c_1}e^-)e^{9'}+\frac{k_1}{4r^2}e^{-2c_1}(e^{12}-e^{34}-e^{56}+e^{78'}) \,.
\ee
We can see that $\Omega$ is really the K\"ahler form for the ten dimensional spatial manifold $\CB$ by redefining the spatial vielbeins
\be
\Omega = \bar{e}^{12}-\bar{e}^{34}-\bar{e}^{56}+\bar{e}^{78'}+\bar{e}^{9'10'} \,.
\ee
where
\be
e^i &=& \D^{-1/2}\bar{e}^i \,, \quad e^{10'} = \D^{-1}\tfrac{1}{2r^2}(c_2 e^+ +2e^{2c_1}e^-) \;,
\nn \\
\D &=& \left(\frac{c_2e^{2c_1}}{r^4}\right)^{1/2} = \frac{k_1e^{-2c_1}}{4r^2} \,.
\ee

\newpage
\section{$S^2\times T^2$ solution}
To date much of our understanding of non-relativistic geometric duals comes from work in type IIB supergravity  \cite{hrr,mmt,abm,hartnoll,bobev,Donos:2009en,Donos:2009xc,Donos:2009zf}. A thorough account of the supersymmetry preserved appeared in \cite{Donos:2009zf}, and a prescription was given therein to construct a special subclass of solutions based on five-dimensional Sasaki-Einstein spaces that realise the $\mathcal{N} = 2$ Super-Schr\"{o}dinger algebra. In this section we consider an explicit example from that class, uplift it to M-theory and comment on the Killing spinors preserved.  

The general form of the solutions presented in \cite{Donos:2009zf} may be written as 
\be
ds^2 &=& -\frac{h dt^2}{r^4} + \frac{2dt d\psi+dr^2+d\vec{x}^2}{r^2}+ds^2_{SE_5}, 
\nn \\
F_5 &=& (1+*_{10})dx^+\wedge dx^- \wedge dx_1 \wedge dx_2 \wedge d(1/r^4), 
\nn \\
G_3 &=& dx^+ \wedge d(\sigma/r^2). 
\ee
Here $ds^2_{SE_5}$ denotes the metric on a five-dimensional Sasaki-Einstein space $SE_5$ and $\sigma$ is a \textit{complex} one-form on the Calabi-Yau cone $CY_3$ dual to a Killing vector on $SE_5$. The function $h$ is given by
\be
h=|\sigma|^2_{SE} + \thalf (\eta_{SE})^\mu L_\mu,
\quad
L \equiv i \Lie_{\sigma^*}^{SE_5}\sigma,
\label{h-def}
\ee 
where $\eta_{SE}$ is the one-form dual to the Reeb Killing vector on $SE_5$ and $L$ is given in terms of the Lie-derivative with respect to the vector dual to $\sigma^*$.

In constructing an explicit example in this class we adopt the $SE_5$ metric discovered in \cite{gaunt3, gaunt6}
\be
ds^2_{SE_5} &=& \frac{1-cy}{6}(d\theta^2+\sin^2\theta d\phi^2)+e^{-6\lambda}\sec^2\zeta dy^2 +\tfrac{1}{9}\cos^2\zeta (D\beta)^2
\nn \\
&& +e^{6\lambda}\left(dz+\frac{ac-2y+cy^2}{6(a-y^2)}D\beta\right)^2
\label{s2t2-metric}
\ee
where
\be
D\beta &=& d\beta -\cos\theta d\phi \,,
\nn \\
e^{6\lambda} &=& \frac{2(a-y^2)}{1-cy} \,,
\nn \\
\cos^2\zeta &=& \frac{a-3y^2+2cy^3}{a-y^2} \,.
\ee

%\paragraph{Uplift formula}

%We take the following Kaluza-Klein ansatz for M-theory on $T^2$:
%\be
%ds^2_M &=& e^{2\l} g_{\m\n}dx^\m dx^\n + e^{-4\l} h_{ab} D\vph^a D\vph^b \,,
%\\
%C_3 &=& C_{(1)} D\vph^1 D\vph^2 + C_{(2)a} D\vph^a + C_{(3)} \,,
%\ee
%where $D\vph^a = d\vph^a +A^a_{(1)} = d\vph^a +A^a_{\m} dx^\m$, $\det(h_{ab}) = 1$
%and the products of forms in the last line are wedge products.

%The corresponding IIB solution take the form
%\be
%ds^2_{IIB} &=& g_{\m\n}dx^\m dx^\n + e^{6\l} (Dz)^2 \,,
%\\
%h_{ab} &=& h_{ab}
%\\
%B_2^a &=& A^a_{(1)} Dz + \e^{ab} C_{(2)b} \,,
%\\
%C_4 &=& C_{(3)} Dz + C_{(4)} \,.
%\ee
%where $Dz = dz+C_{(1)}$.

In general preserving six supersymmetries requires a judicious choice for $\sigma$. As explained in \cite{Donos:2009zf}, one requires $\sigma$ is chosen so that its exterior derivative on $CY_3$, $d(\tfrac{\sigma}{r^2})$, is of type (1,1) and primitive\footnote{In the earlier non-supersymmetric solutions \cite{hrr,mmt,abm} $\sigma$ was chosen dual to the Reeb vector meaning that the exterior derivative was proportional to the K\"{a}hler two-form.}.  Within these constraints, we choose $\sigma$ to be the one-form dual to the Killing vector $V_\sigma = \kappa_1\partial_\phi-\tfrac{\kappa_2}{6}\partial_z$, which is the sum of two Cartans in three on (\ref{s2t2-metric}) and $\kappa_1$ and $\kappa_2$ are arbitrary complex constants. For this choice, $L$ is zero and $h$ follows from (\ref{h-def}), $h = |\sigma|_{SE}^2$. 

By T-dualising to type IIA and uplifting this solution, we obtain a deformation of the class of warped supersymmetric $AdS_5 \times M_6$ solutions with base space $M_4 = S^2 \times T^2$, originally discovered in \cite{gaunt3}. The explicit solution has manifest Schr\"{o}dinger symmetry and may be expressed as follows 
\be
ds^2_{11D} &=& e^{2\lambda}\left(-\frac{\tilde{h} dt^2}{r^4} + \frac{2dt D\psi+dr^2+d\vec{x}^2}{r^2}\right) + e^{2\lambda}ds^2_{M_6} \,,
\nn \\
ds^2_{M_6} &=&\frac{1-cy}{6}(d\theta^2+\sin^2\theta d\phi^2)+e^{-6\lambda}\sec^2\zeta dy^2+\tfrac{1}{9}\cos^2\zeta (D\beta)^2+e^{-6\lambda}(d\vph_1^2+d\vph_2^2) \,,
\nn \\
F_4 &=& -\tfrac{2}{9}(1-cy)dy\wedge D\beta \wedge Vol(S^2) +d\left(\frac{ac-2y+cy^2}{6(a-y^2)}D\beta \wedge d\vph_1 \wedge d\vph_2 \right)
\nn \\
&&+d\left(\frac{\tilde{\sigma}}{r^2}\right)\wedge dt \wedge (Im(\kappa_1) d\vph_1 - Re(\kappa_1) d\vph_2) \,,
\label{s2t2-sol}
\ee
where 
\be
\tilde{h} &=& |\kappa_1|^2 \tilde{\sigma}^2_{M_6} \,,
\nn \\
D\psi &=& d\psi -Re(A) d\vph_1 -Im(A) d\vph_2 \,,
\nn \\
A &=& \kappa_1 \frac{ac-2y-cy^2}{6(a-y^2)}\cos\theta + \tfrac{1}{6}\kappa_2 \,,
\nn \\
\tilde{\sigma} &=& -\tfrac{1}{9}\cos^2\zeta\cos\theta D\beta +\frac{1-cy}{6}\sin^2\theta d\phi \,.
\ee
Here $\tilde\sigma$ is the dual one form of $\partial_\phi$ with respect to the metric on $M_6$, and one may check that when $\kappa_1 = \kappa_2 = 0$, this reduces to the original undeformed solutions \cite{gaunt3}.

This explicit example (\ref{s2t2-sol}) is supersymmetric, admitting six Killing spinors: two kinematical, two Poincar\'e and two superconformal Killing spinors. We now turn to detailing how it preserves these supersymmetries and what form the Killing spinors take. Since these solutions are deformations of solutions of \cite{gaunt3} with parameters $\kappa_1, \kappa_2$, we can incorporate some of the expressions from \cite{gaunt3} wholesale. 

In calculating the Killing spinors we can write the 11D gamma matrices as
\be
\G^a &=& \rho^a \otimes \gamma_7 \,,
\nn \\
\G^m &=& {\bf 1} \otimes \gamma^m \,,
\nn \\
\gamma_7 &\equiv& \gamma_1\dots\gamma_6 \,.
\ee
where $a, b = +,-,1,2,3$ and $m, n=1,2, \dots, 6$ are indices on $Sch_5$ and $M_6$ respectively. Here we take the vielbein as
\be
e^+ &=& \frac{e^{\lambda}}{r^2}dt \,, \quad e^- = e^{\lambda}\left(-\frac{h}{2r^2}dt+D\psi\right) \,,
\nn \\
e^1 &=& \frac{e^{\lambda}}{r}dx^1 \,, \quad e^2 = \frac{e^\lambda}{r}dx^2 \,, \quad e^3 = \frac{e^{\lambda}}{r}dr \,,
\nn \\
e^4 &=& e^{c_1}(\frac{1-cy}{6})^{1/2}\sigma_1 \,, \quad e^5 = e^{c_1}(\frac{1-cy}{6})^{1/2}\sigma_2 \,,
\nn \\
e^6 &=& e^{-2\lambda}\sec\zeta dy \,, \quad e^7 = \frac{e^\lambda}{3}\cos\zeta D\beta \,,
\nn \\
e^8 &=& e^{-2\lambda}d\vph_1 \,, \quad  e^9 = e^{-2\lambda}d\vph_2 \,.
\ee
The dynamical Killing spinors may then be written as 
\be
\epsilon_d &=& \frac{\kappa_1}{4r}\left[-i\tfrac{2}{3}\cos\zeta \cos\theta\G^{+78}+\sqrt{\tfrac{2}{3}(1-cy)}e^{i\beta}\sin\theta\G^{+58}\right]\eta+\tfrac{1}{r} \eta
\ee
and $\eta$ is the product $\psi\otimes e^{\lambda/2}\xi$, with $\psi$ denoting the $AdS_5$ Killing spinors
\be
\nabla_a \psi = \tfrac{i}{2}\rho_a \psi \,,
\label{s2t2-spinor1}
\ee
and $\xi$ being further decomposed in terms of two orthogonal unit-norm chiral spinor $\eta_i$ \cite{gaunt3} 
\be
\xi &=& \sqrt{2}\cos\alpha\, \eta_1 +\sqrt{2}\sin\alpha\, \eta_2^*
\ee
where $\cos2\alpha = \sin\zeta$. These two spinors satisfy the following projection conditions 
\be
&&\gamma^{12}\eta_1 = -\gamma^{34}\eta_1=\gamma^{56}\eta_1 = i\eta_1 \,,
\nn \\
&&\gamma^{12}\eta_2 = \gamma^{34}\eta_2=\gamma^{56}\eta_2 = -i\eta_2 \,,
\nn \\
&&\gamma^3\eta_2^* = \eta_1 \,.
\ee
The original geometries preserve eight Killing spinors. In the presence of the deformation to bring the geometry to a Schr\"{o}dinger invariant form, 
we discover the additional projection conditions
\be
\rho^3\psi &=& i\psi \,,
\nn \\
\Gamma^-\eta &=& 0 \,.
\ee
With these additional constraints, the spinor $\epsilon_d$ satisfies the Killing spinor equations. The kinematical, $\epsilon_k$, and superconformal, $\epsilon_s$, Killing spinors can then be constructed from the algebra as was illustrated in the earlier text
\be
\epsilon_k &=& \tfrac{1}{\sqrt{2}}\Gamma^{+1}\eta \,,
\nn \\
\epsilon_s &=& \left[t-\thalf r\G^+(x_i\G^i+r\G^3)\right]\epsilon_d \,.
\ee

\section{Discussion}

%\paragraph{Non-existence of spectator supercharges}

We saw in section 2 that the anti-commutations of two spectator supercharges give the generators for $SU(2)_1\times SU(2)_2$ as well as the central element $M$.
\be
\left\{ \bar{q}^{a\dot{a}} , q_{b\dot{b}} \right\}
= \half \d^a_b \d^{\dot{a}}_{\dot{b}} M 
- \d^a_b R^{\dot{a}}{}_{\dot{b}} +\d^{\dot{a}}_{\dot{b}} R^a{}_b \,,
\label{spectator-algebra}
\ee
From the geometric point of view, 
the spinor bi-linears $\bar{\e} \G^m \e$ made of the Killing spinors 
$\e_{a\dot{a}}$ corresponding to $q_{a\dot{a}}$ should produce the Killing vectors 
for the generators on the right hand side of 
(\ref{spectator-algebra}). 
Now, recall that $q_{a\da}$ commute with $(H,D,C,P,\bar{P},M)$. 
Inspecting the spinorial Lie derivatives (\ref{spinor-Lie}), especially $\Lie_C\e$, 
we find that $\e_{a\dot{a}}$ must be annihilated by $\G^+$. 
This implies that all bi-linears constructed from $\e_{a\dot{a}}$ 
can have non-zero components only in the $(x^-)$-directions 
much like the kinematical supercharges $(q,\bar{q})$ discussed earlier:
\be
\bar{\e}^{a\dot{a}}\G^{m}\e_{b\dot{b}} = 0 \quad (\mbox{except for }m=-)\,.
\ee
In particular, the generators for $SU(2)_1\times SU(2)_2$ symmetry cannot be produced by the Killing spinors. We thus proved without much computation that 
the Killing spinors for the spectator supercharges with desired algebraic property 
do not exist within our ansatz. 

Even if we give up the $SU(2)_1\times SU(2)_2$ generators in (\ref{spectator-algebra}), 
it is still impossible to obtain eight extra Killing spinors 
as one can see from the following counting argument. 
We argued above for the projection condition $\G^+ \e_{a\dot{a}}=0$. 
The fact that $q_{a\dot{a}}$ transform in the same way 
under the two $SU(2)$ groups imply that $\p_v \e_{a\dot{a}}=0$, which 
together with $\Lie_M \e_{a\dot{a}}=0$ 
yield another projection condition, $
\G^{3456}\e_{a\dot{a}} = -\e_{a\dot{a}}$.
Finally, since $\e_{a\dot{a}}$ are null Killing spinors, 
the results of \cite{gaunt2} enforces yet another condition, $
\G^9 \e_{a\dot{a}} = \e_{a\dot{a}}$.
Three mutually orthogonal projection conditions leave 
at most $32/2^3=4$ independent components, 
so the possiblity of eight extra spinors is excluded.  
%This result is not consistent with the previous assumption that we might have eight spectator Killing spinors.

We have shown that a supergravity background dual to the NR-ABJM theory 
preserving the super-Schr\"odinger symmetry and all the global symmetries 
does not exist. We do not have a clear physical understanding of why 
this is the case. We end this paper with two possible directions 
we may pursue to find an explanation.
\footnote{We thank Seok Kim for discussions on the second possibility.} 

First, it is conceivable that the singularity problem of the unpolarized BW/LLM solution 
mentioned in section 2 is unavoidable, so that even if we 
find a good way to take the non-relativistic limit, the resulting 
geometry would be necessarily singular. If this is true, 
we may need to doubt either the existence of the NR-ABJM theory as a quantum field theory or the validity of non-relativistic holography. 

Second,
note that we have searched for a gravity solution preserving 
all Schr\"odinger and global symmetries apart from the non-zero particle number 
($M$-eigenvalue). Via holography, it would correspond to 
a ground state of the NR-ABJM theory for a fixed non-zero particle number 
that preserves all the symmetries. It is not obvious a priori 
whether such a ground state should exist in the field theory. 
If holography works, the non-existence of the fully symmetric gravity solution may 
be an indication that the ground states of the field theory 
necessarily break some parts of the symmetries. 
It would be interesting to test this idea by studying the spectrum of the field theory directly.

\vspace{1cm}

\section*{Acknowledgments}
We are grateful for Oren Bergman, Seok Kim, Ki-Myeong Lee and Sungjay Lee for discussions.
SL is grateful to the Aspen Center for Physics for hospitality,
where parts of this work were carried out.
The work of JJ, HK and SL is supported in part by the
National Research Foundation of Korea (NRF) Grants No. 2007-331-C00073, 2009-0072755 and 2009-0084601. 
The work of SL is also supported in part by the NRF Grant No. 2005-0049409 
through the Center for Quantum Spacetime (CQUeST) of Sogang University.

%\vskip 2cm 

\newpage

\centerline{\large \bf Appendix}

\appendix

\section{Notations and Conventions \label{conv}}

\paragraph{11-dimensional supergravity}

The bosonic part of the Lagrangian is
\be
2\k_{11}^2 \CL =
 R \,\ast 1- \half F \wedge \ast F
-\frac{1}{6} A \wedge F \wedge F.
\ee
The fermionic part of the SUSY transformation rule becomes
\be
\d\psi_M = \grad_M\e +\frac{1}{12\cdot 4!}F_{IJKL}
\left({F^{IJKL}}_M - 8 \d^{I}_{M} \G^{JKL} \right)\e.
\ee

\paragraph{Euler-angle coordinates}

We take the metric of the $\IR^4$ to be
\be
ds^2 = dr^2 + \frac{r^2}{4}(\s_1^2+\s_2^2+\s_3^2) \,,
\ee
where the left-invariant one-forms are defined in terms of Euler angle coordinates by
\be
\s_1 &=& +\sin\psi\, d\th + \cos\psi \sin\th\,d\phi \,,
\nn \\
\s_2 &=& -\cos\psi\, d\th + \sin\psi \sin\th\,d\phi \,,
\nn \\
\s_3 &=&  d\psi-\cos\th \,d\phi \,.
\ee
The $SU(2)_L$ action is generated by the Killing vectors,
\be
V_1 &=& +\sin\phi\, \p_\th + \cot\th \cos\phi\,\p_\phi +\csc\th \cos\phi\,\p_\psi\,,
\nn \\
V_2 &=& -\cos\phi\, \p_\th + \cot\th \sin\phi\,\p_\phi +\csc\th \sin\phi\,\p_\psi\,,
\nn \\
V_3 &=&  -\p_\phi \,,
\ee
while the $SU(2)_R$ action is generated by
\be
\widehat{V}_1 &=& -\sin\psi\, \p_\th - \cot\th \cos\psi\,\p_\psi -\csc\th \cos\psi\,\p_\phi\,,
\nn \\
\widehat{V}_2 &=& +\cos\psi\, \p_\th - \cot\th \sin\psi\,\p_\psi -\csc\th \sin\psi\,\p_\phi\,,
\nn \\
\widehat{V}_3 &=&  -\p_\psi \,.
\ee
They satisfy the following relations,
\be
&& [V_A ,V_B] = \e_{ABC} V_C \,,
\;\;\;\;\;
[\widehat{V}_A ,\widehat{V}_B] = \e_{ABC} \widehat{V}_C \,,
\nn \\
&&d\s_A = \thalf \e_{ABC} \s_B \wedge \s_C \,,
\;\;\;\;\;
\CL_{V_A} \s_B = 0 \,,
\;\;\;\;\; \CL_{\widehat{V}_A} \s_B = \e_{ABC} \s_C \,.
\ee

%\newpage

\section{Bena-Warner/Lin-Lunin-Maldacena solution \label{bw-llm}}

\paragraph{Review of LLM}
The geometry is specified by a function $z(x,y)$ defined on the upper half plane
$(y\ge0)$. The function $z$ satisfies the differential equation
\be
\p_x^2 z + y \p_y(y^{-1} \p_yz) = 0 \,,
\ee
with a boundary condition at $y=0$. Regularity of the geometry requires
that $z=\pm 1/2$ on the boundary. It is useful to introduce a few
additional variables,
\be
V\; &:& \p_x z = -y \p_y V \,, \;\; \p_y z = y \p_x V \,,
\\
G\; &:& z = \thalf \tanh G \,,
\\
h \;&:& h^{-2} = 2y \cosh G \,, \label{hG}
\\
H \;&:& H = h^2 - h^{-2}V^2 \,.
\ee
In terms of these variables,
the most general supergravity solution with sixteen supercharges
and $SO(1,2)\times SO(4)\times SO(4)$ isometry can be written as
\be
ds^2 &=& H^{-2/3} (-dt^2+dw_1^2+dw_2^2) + H^{1/3}\left[ h^2(dy^2+dx^2)+ye^G d\Om_3^2+ye^{-G} d\tilde{\Om}_3^2 \right] \,,
\nn \\
F &=& -d(H^{-1}h^{-2}V) \wedge dt \wedge dw_1 \wedge dw_2
\nn \\
&&-\textstyle{\frac{1}{4}} H \left[ e^{-3G}*_2 d(y^2e^{2G}) \wedge d\tilde{\Om}_3 +e^{3G}*_2 d(y^2e^{-2G}) \wedge d\Om_3 \right] \,.
\ee
Here, $*_2$ is the flat epsilon symbol in the $(x,y)$ plane.

\paragraph{Mass deformed AdS$_4$ (without polarization)}
Section 4.2 of Bena-Warner gives a solution
describing the flux deformation of AdS$_4\times S^7$.
It is instructive to rewrite the solution in the LLM coordinates.
How to translate between the two coordinates is explained below.

The result is
\be
\label{z00}
z &=& \frac{x}{2\sqrt{x^2+y^2}} \left[ 1 - \frac{3\g^2y^2}{(x^2+y^2)^2}\right] \,,
\\
V &=& \frac{1}{2\sqrt{x^2+y^2}}  \left[ 1 + \frac{\g^2(2x^2-y^2)}{(x^2+y^2)^2}\right] \,,
\\
h^2 &=& \frac{1}{2\sqrt{x^2+y^2}} \left[ 1+\frac{6\g^2x^2}{(x^2+y^2)^2}
-\frac{9\g^4x^2y^2}{(x^2+y^2)^4}\right]^{1/2} \,,
\\
H &=& \frac{\g^2}{(x^2+y^2)^{3/2}}
\left[ 1-\frac{\g^2(4x^2+y^2)}{2(x^2+y^2)^2}\right]
\left[ 1+\frac{6\g^2x^2}{(x^2+y^2)^2}
-\frac{9\g^4x^2y^2}{(x^2+y^2)^4}\right]^{-1/2} \,.
\label{H00}
\ee
%The parameter $\g$ can be rescaled to $1$ by using the
%scale symmetry $(x,y)\goto (\l x, \l y)$ under which
%the functions transform as $(z,V,h^2,H) \goto (z,\l^{-1}V,\l^{-1}h^2, \l^{-3}H)$.
The parameter $\g$ is related to those of BW \cite{BW} by
\be
\g^2 = (128 L^6 \b^2 R^6 )_{BW} \,.
\ee

In the UV region $(x,y \gg \g)$, all the square brackets in (\ref{z00})-(\ref{H00}) can be ignored, and
%Going back to the Bena-Warner coordinates,
%\be
%x = \thalf(u^2-v^2)\,, \;\;\; y=uv \,,
%\ee
we recover the AdS$_4 \times S^7$ geometry upon a suitable constant rescaling
of coordinates.

% ($r^2\equiv u^2+v^2$):
%\be
%ds^2 &=& \frac{r^4}{R^4} (-dt^2+dw_1^2+dw_2^2) + \frac{R^2}{r^2} \left[ du^2+dv^2 +u^2 d\O_3^2+v^2 d\tilde{\O}_3^2 \right] \,,
%\\
%F_4 &=&  \mbox{{\bf we may need to introduce explicit factors of $\g$ in (2.6)}}\,.
%\ee

%\subsection{Geodesic}

%There exists a null geodesic at $(x,y)=(0,\gamma)$. The derivation is as follows.
%Let us examine a family of null geodesics with $x=0$ and all angles but $\psi$ (the diagonal U(1); see section 3 for the definition of $\psi_+$) set to constant.

%\be
%E &\equiv& g_{tt} \frac{dt}{ds} = H^{-2/3} \frac{dt}{ds}
%\\
%J &\equiv& g_{\psi\psi} \frac{d\psi}{ds} = \half H^{1/3} (2y \cosh G) \frac{d\psi}{ds}
%\ee

%\be
%g_{\m\n} \frac{dx^\m}{ds} \frac{dx^\n}{ds} = 0
%\ee

%\be
%- H^{2/3} E^2 + ( H^{1/3} y \cosh G)^{-1} J^2  + H^{1/3} h^2 \dot{y}^2 = 0 \,.
%\ee
%In other words,
%\be
%H^{-1/3} h^2 \dot{y}^2 + (H y \cosh G)^{-1} J^2 = E^2
%\ee

%\be
%H  = \frac{R^6}{8y^3} \left( 1-\frac{\g^2}{2y^2} \right)
%\ee

%\be
%(H y \cosh G)^{-1} = \frac{8 y^2}{R^6}
%\ee

%\subsection{BMN-like limit}

\paragraph{Relation between Bena-Warner and Lin-Lunin-Maldacena}

We compare the notations of Bena and Warner (BW) \cite{BW} and those of
Lin, Lunin and Maldacena (LLM) \cite{LLM}. This was already done in
appendix C of \cite{LLM}, but it contained some minor errors.

The BW and LLM metrics read,
\be
ds^2_{BW} &=& 16L^4 e^{2B_0}(-dt^2+dw_1^2+dw_2^2) + e^{2B_1-B_0}(du^2+dv^2)
\nn \\
&& +u^2e^{2B_3-B_0} d\Omega_3^2 + v^2 e^{-2B_3-B_0} d\widetilde{\Omega}_3^2 \,,
\\
ds^2_{LLM} &=& H^{-2/3} (-dt^2+dw_1^2+dw_2^2)
\nn \\
&&+ H^{1/3}\left[ h^2(dy^2+dx^2)+ye^G d\Om_3^2+ye^{-G} d\tilde{\Om}_3^2 \right] \,,
\ee
which lead to the identifications,
\be
&&H^{-2/3} = 16L^4 e^{2B_0} \,, \;\;\;
ye^G = 4L^2 u^2 e^{2B_3} \,, \;\;\;
ye^{-G} = 4L^2 e^{-2B_3} \,,
\nn \\
&&h^2(dx^2+dy^2) = 4L^2 e^{2B_1} (du^2+dv^2) \,.
\label{iden-all}
\ee
Combining the $G$-$B_3$ relations, we find
\be
e^G = \frac{u}{v} e^{2B_3} \,, \;\;\; y= 4L^2 uv \,.
\label{Gy}
\ee
Orthogonality of the coordinates implies
\be
x = 2L^2 (u^2-v^2) \,.
\label{x-uv}
\ee
Putting (\ref{Gy}) and (\ref{x-uv}) back to (\ref{iden-all}), we find
\be
h^{-2} = 4L^2 e^{-2B_1} (u^2+v^2) \,.
\label{h-uv}
\ee
As a cross check, we note that inserting (\ref{Gy}) and (\ref{h-uv})
into the LLM relation (\ref{hG}),
\[
h^{-2} = y(e^G+e^{-G}) \,,
\]
reproduces eq. (43) of \cite{BW},
\[
e^{-2B_1} (u^2+v^2) = u^2 e^{2B_3} + v^2 e^{-2B_3} \,.
\]

In translating the BW solution into the LLM form,
it is most convenient to use first
\be
2 z= \frac{e^{2G}-1}{e^{2G}+1}
=  \frac{u^2 e^{4B_3} - v^2}{u^2 e^{4B_3} + v^2} \,,
\ee
and then use the LLM formulas to compute other quantities.

%\vskip 2cm 

\end{document}